\documentclass[aps,prd,floatfix,nofootinbib,superscriptaddress,preprint]{revtex4-1}
\usepackage{amstext,amssymb}
\usepackage{graphicx}
\usepackage{color}
\usepackage{epsfig}
\usepackage{xspace}
\usepackage{color}
\usepackage{units}
\usepackage{slashed} % feynman slash notation via \slashed{p}

\DeclareMathSymbol{\mlq}{\mathord}{operators}{``}
\DeclareMathSymbol{\mrq}{\mathord}{operators}{`'}

\begin{document}
\title{Alternative formulation of left-right symmetry\\ with $B-L$ conservation and purely Dirac neutrinos}
\author{Patrick D. Bolton}
\email{patrick.bolton.17@ucl.ac.uk}
\affiliation{Department of Physics and Astronomy, University College London, Gower Street, London WC1E 6BT, United Kingdom}
\author{Frank F. Deppisch}
\email{f.deppisch@ucl.ac.uk}
\affiliation{Department of Physics and Astronomy, University College London, Gower Street, London WC1E 6BT, United Kingdom}
\author{Chandan Hati}
\email{chandan.hati@clermont.in2p3.fr} 
\affiliation{Laboratoire de Physique de Clermont CNRS/IN2P3 -- UMR 6533,
 4 Avenue Blaise Pascal, 63178 Aubi\`ere Cedex, France}
\author{Sudhanwa Patra}
\email{sudhanwa@iitbhilai.ac.in}
\affiliation{Indian Institute of Technology Bhilai, Raipur 492015, Chhattisgarh, India}
\author{Utpal Sarkar}
\email{utpal@phy.iitkgp.ernet.in}
\affiliation{Department of Physics, Indian Institute of Technology Kharagpur, Kharagpur 721302, India}

\begin{abstract}
We propose an alternative formulation of a Left-Right Symmetric Model (LRSM) where the difference between baryon number ($B$) and lepton number ($L$) remains an unbroken symmetry. This is unlike the conventional formulation, where $B-L$ is promoted to a local symmetry and is broken explicitly in order to generate Majorana neutrino masses. In our case $B-L$ remains a global symmetry after the left-right symmetry breaking, allowing only Dirac mass terms for neutrinos. In addition to parity restoration at some high scale, this formulation provides a natural framework to explain $B-L$ as an anomaly-free global symmetry of the Standard Model and the non-observation of $(B-L)$-violating processes. Neutrino masses are purely Dirac type and are generated either through a two loop radiative mechanism or by implementing a Dirac seesaw mechanism.
\end{abstract}

\maketitle

\section{Introduction} 
With the discovery of the Higgs boson the last missing piece of evidence 
confirming the Standard Model (SM) of particle physics has been obtained. However, the observation 
of neutrino oscillations has established non-vanishing neutrino masses, which is undeniable 
evidence of physics beyond the SM. In the SM the left-handed fermions transform as electroweak doublets while 
the right-handed fermions transform as singlets due to parity violation. Thus, it is natural to look 
for a left-right symmetric theory at a high energy scale, where both the left-handed and the right-handed fermions transform on an equal footing under the gauge group and parity is restored. At some high energy the left-right symmetric gauge group and parity are broken spontaneously, which explains the observed parity violation at low energies. 

This motivates the left-right symmetric model (LRSM), in which the SM gauge group $SU(3)_c \times SU(2)_L \times U(1)_Y$ is extended to make it left-right symmetric $SU(3)_c \times SU(2)_L \times SU(2)_R \times U(1)_X$ \cite{lr}. Several versions of the LRSM exist in the literature (see for example \cite{Hati:2018tge} for a recent review) and in almost all of these models one identifies the generator of the group $U(1)_X$ with the $B-L$ symmetry, where $B$ is the baryon number  and $L$ is the lepton number\footnote{For some exceptions see e.g. Refs.~\cite{London:1986dk, Dhuria:2015hta}.}. For the SM particles this identification follows simply from the charge equation relating the SM gauge group to the LRSM gauge group which is broken by the conventional choice of a triplet Higgs scalar. If this choice is generalised for the right-handed neutrinos one can then generate small Majorana neutrino masses for the neutrinos through the seesaw mechanism \cite{seesaw}. However, in general this choice is not unique for new fermions added to the SM spectrum or for alternative Higgs sectors.

In the conventional LRSM, at some high energy scale compared to the electroweak symmetry breaking scale the left-right symmetric gauge symmetry group can be written as
\begin{equation}{\label{1.1}}
	 G_{LR} \equiv SU(3)_c \times SU(2)_L \times SU(2)_R \times U(1)_X\,,
 \end{equation}
 which breaks down to the SM gauge group $SU(3)_c \times SU(2)_L \times U(1)_Y$. The electric charge is related to the generators of the gauge groups by the relation
\begin{equation}\label{1.2}
    Q = T_{3L} + T_{3R} + \frac{X} { 2} = T_{3L} + Y\,.
\end{equation}
In the conventional case the quantum number $X$ is identified with the $B-L$ symmetry, so that 
$B-L$ becomes a local gauge symmetry of the model. Consequently, the left-right symmetry breaking can 
induce several $(B-L)$-violating interactions, including the generation of Majorana neutrino masses 
via a seesaw mechanism. The transformations of the left- and right-handed fermions under the left-right symmetric gauge group $G_{LR} \equiv SU(3)_c \times SU(2)_L \times SU(2)_R \times U(1)_{B-L}$ are given by
\begin{eqnarray}\label{1.3}
q_L =\pmatrix{ u_L \cr d_L }  \equiv [ 3,2,1, {1 \over 3} ]\,,~~&& q_R
=\pmatrix{ u_R \cr d_R }  \equiv [ 3,1,2, {1 \over 3}]\,,
\nonumber \\
\ell_L  =\pmatrix{ \nu_L \cr e_L }  \equiv [ 1,2,1, {-1  } ]\,,~~&&
\ell_R  =\pmatrix{ \nu_R \cr e_R }  \equiv [ 1,1,2, -1]\,.
\end{eqnarray}
Left-right symmetry naturally includes the right-handed neutrinos $\nu_R$. The symmetry breaking pattern is given by
\begin{eqnarray}\label{1.4}
&&SU(3)_c \times SU(2)_L \times SU(2)_R \times U(1)_{X} ~~[
G_{LR}] \nonumber \\
& \stackrel{M_R}{\rightarrow} &SU(3)_c \times
SU(2)_L \times U(1)_Y ~~~~~~~~~~~~~~~~\,[ G_{SM}] \nonumber \\
&\stackrel{m_W}{\rightarrow} &SU(3)_c \times U(1)_Q ~~~
~~~~~~~~~~~~~~~~~~~~~~~~~~~[ G_{em}]\,,  \nonumber
\end{eqnarray}
where $M_R$ corresponds to the $SU(2)_R$ breaking scale. The relevant scalar sector is given by
\begin{eqnarray}
\Phi &=& \pmatrix{ \Phi^{0}_{1} & \Phi^{+}_{1} \cr \Phi^{-}_{2} & \Phi^{0}_{2} }  : [1,2,2,0]\,,\nonumber\\
\Delta_{L} &=&  \pmatrix{   \frac{\Delta^{+}_L}{\sqrt{2}} & \Delta^{++}_L \cr \Delta^{0}_L & -\frac{\Delta^{+}_L}{\sqrt{2}} }_{L} : [1,3,1,2]\,,\nonumber\\
 \Delta_{R}&= & \pmatrix{   \frac{\Delta^{+}_R}{\sqrt{2}} & \Delta^{++}_R \cr \Delta^{0}_R & -\frac{\Delta^{+}_R}{\sqrt{2}}}_{R} : [1,1,3,2]\,.
\end{eqnarray}.
In the conventional LRSM, the $X=B-L$ symmetry is broken by the triplet Higgs scalar $\Delta_R \equiv [1,1,3,2]$ and from left-right parity symmetry one must also have another triplet Higgs scalar 
$\Delta_L \equiv [1,3,1,2]$. For both the triplets $\Delta_{L,R}$, the $U(1)$ quantum number 
is $B-L=-2$. In the absence of any additional symmetry, the gauge symmetry allows 
the interactions of the Higgs triplets with the fermions
\begin{equation}\label{1.5}
{\cal L} = f \ell_L^T C^{-1} \ell_L \Delta_L + f \ell_R^T
C^{-1} \ell_R \Delta_R \,,
\end{equation}
which determine the $B-L$ quantum number of $\Delta_{L,R}$ 
uniquely, allowing the identification $X=B-L$. The SM Higgs doublet breaking the electroweak symmetry 
also gives masses to the fermions, which in the presence of both left- and right-handed fermions 
transforming as doublets dictate that the SM Higgs doublet should be a bi-doublet under 
the group $G_{LR}$:
\begin{eqnarray}\label{1.6}
\phi &\equiv& [1,2,2,0]\,,
\end{eqnarray}
with $X=B-L=0$.

When this conventional model is embedded in grand unified theories like $SO(10)$ GUT, the theory 
contains diquarks ($\Delta_{qq}$ that couple to two quarks) or leptoquarks ($\Delta_{lq}$ that 
couple to a quark and a lepton or an anti-lepton) \cite{dn}. All these scalar fields belong to one 
126-dimensional representation of SO(10) and their quantum numbers are determined by the quantum 
numbers of the fermions, which dictate $X=B-L$. In this conventional formalism there are many sources of $B-L$ 
violation, all of which could affect the lepton asymmetry of the universe, and hence, the baryon 
asymmetry of the universe. Usually one considers mainly the interactions of the right-handed 
neutrinos $N_R$ when studying leptogenesis \cite{lepto1,lepto2} and assumes that all other interactions to decouple before 
$T \approx M_N$, where $M_N$ is the mass of the lightest right-handed neutrino. The lepton asymmetry 
generated by the decays of the lightest right-handed neutrino would then get converted to a baryon 
asymmetry of the universe in the presence of the sphalerons before the electroweak phase transition. 
However, after the decays of the right-handed neutrinos there could be fast $(B-L)$-violating interactions 
originating from the spontaneous breaking of the gauged $B-L$ symmetry \cite{Deppisch:2013jxa, Deppisch:2015yqa, Deppisch:2017ecm, Frere:2008ct, Dev:2015vra, Dev:2014iva, Dhuria:2015wwa, Dhuria:2015cfa}\footnote{For a recent review with some relevant discussion see for example Ref.~\cite{Chun:2017spz}.}. 

A complete study should thus
address all the following interactions: (i) Interactions of the gauge boson $W_R$ with the right-handed leptons, and also with the Higgs triplet $\Delta_R$ which violate $B-L$ quantum numbers \cite{Keung:1983uu, Deppisch:2013jxa, Deppisch:2015yqa, Deppisch:2017ecm, Frere:2008ct, Dev:2015vra, Dev:2014iva, Dhuria:2015wwa, Dhuria:2015cfa}. In some models, these interactions can also generate a lepton asymmetry. (ii) Interactions of the diquark Higgs scalars $\Delta_{qq}$ with themselves and with the dilepton Higgs scalars \cite{Mohapatra:1980de,Pati:1983zp,Babu:2012iv,Phillips:2014fgb,Hati:2018cqp}. When a model predicts neutron-antineutron oscillation, light diquark Higgs scalars are predicted. These models may wash out the lepton asymmetry generated by the right-handed neutrino decays. (iii) The interactions of the right-handed triplet Higgs scalars  $\Delta_R$ \cite{Bambhaniya:2015wna, Dutta:2014dba, Dev:2016dja, Mitra:2016wpr} can also affect the lepton asymmetry generated by other mechanisms. (iv) The left-handed triplet Higgs scalars $\Delta_L$ can generate a lepton asymmetry and also a neutrino mass, after the right-handed neutrinos decay \cite{Ma:1998dx,Magg:1980ut,Cheng:1980qt,Lazarides:1980nt,Lazarides:1998iq}. Even when $M_\Delta > M_N$, the Higgs decay can generate an asymmetry, which is not affected by the slow lepton number violating decays of the right-handed neutrinos.

In what follows, we will construct a formulation of an LRSM with an unbroken $B-L$ symmetry, where all these interactions are absent because $B-L$ is not spontaneously broken and consequently one ends up with very different phenomenology and signatures.  First, we note that in general, one can define a new quantum number $\zeta$, such that in Eq. (\ref{1.1}) we have,
\begin{equation}\label{1.7}
X = (B-L) + \zeta\,.
\end{equation}
Thus, if $\zeta \neq 0$, then $B-L$ can also become a global unbroken symmetry, independent of the left-right symmetry. In this work we point out an alternative scheme of left-right symmetry breaking, where $B-L$ is no longer considered to be a local gauge symmetry, but remains an unbroken symmetry. Consequently, all fermions including the neutrinos are Dirac particles. Interestingly, in this model the neutrinos can have tiny Dirac masses generated through either a two loop radiative correction or a Dirac seesaw mechanism \cite{fer-mass} depending on the Higgs sector of the model. This formulation also provides a natural framework to explain $B-L$ as a global symmetry of the SM and can explain the non-observation of any $(B-L)$-violating processes. The baryon asymmetry of the universe can be explained in this formulation through $(B-L)$-conserving neutrinogenesis mechanism \cite{dl1,dl2,Gu:2007mc}.

We would like to emphasise that historically, the original formulation of the LRSM \cite{lr} entertained the possibility of purely Dirac masses for neutrinos as explored in Ref.~\cite{Branco:1978bz}, however due to the breaking of local $B-L$ symmetry together with $SU(2)_R$, the Dirac mass of neutrinos were susceptible to corrections due to Majorana contributions induced by the $B-L$ violating operators in a UV complete theory. This subsequently led to the realisation of Majorana neutrino masses by introducing seesaw mechanism e.g. using triplet Higgs, which is one of the most interesting aspects of these formalisms. On the other hand, in our formalism we explore a potential alternative to the above formalism where the pure Dirac nature of neutrino masses is protected by the unbroken $B-L$ global symmetry, forbidding any possibility of any dimension five or higher lepton number violating operators. Furthermore, the smallness of the neutrino masses is naturally ensured by a two-loop radiative contribution in one of the variants where the tree level and one loop contributions are forbidden by the construction of the model. In another variant of the model the smallness of the Dirac neutrino masses are realised by a Dirac seesaw mechanism and due to the unbroken $B-L$ global symmetry such small Dirac masses are protected against any new Majorana corrections.

The plan for rest of the paper is as follows. In Section~\ref{sec2}, we present an LRSM with an unbroken $B-L$ symmetry where no Higgs bi-doublet is present and the Higgs sector consists of a right-handed doublet, a left-handed doublet  and a parity-odd singlet. In this scenario the quark masses and the charged lepton masses are generated through a seesaw mechanism introducing new vector-like states, while the neutrino masses are generated radiatively at the two loop level. In Section~\ref{sec:para}, we study the two loop radiative contribution in the context of neutrino masses and mixing by constructing a left-right symmetric parametrisation \'{a} la Casas-Ibarra and present a phenomenological numerical analysis for a minimal $2 \times 2$ case, showing the dependence of the PMNS mixing matrix angle on the hierarchy of heavy charged lepton masses and the left-right symmetry breaking scale. In Section~\ref{app:0}, we present another alternative realisation of an LRSM with a global $B-L$ symmetry in the presence of a bi-doublet Higgs. In this scenario the quarks acquire their masses through the vacuum expectation value of the bi-doublet, while the charged and the neutral lepton masses are generated through Dirac seesaw mechanism in the presence of heavy vector-like states. In Section~\ref{sec:pheno}, we outline the observable phenomenology of this formulation and discuss constraints from various considerations. Finally in Section~\ref{sec:conclusion} we conclude and comment on the possible implementation of a dark matter candidate and leptogenesis mechanisms to generate the observed baryon asymmetry of the universe in this scenario.

\section{Left-Right Symmetric Model with an unbroken $B-L$ symmetry}{\label{sec2}}
The fermion content of this model is the same as that given in Eq.~(\ref{1.3}). In addition we will 
add vector-like fermions. For the left-right symmetry breaking, we now use a doublet 
Higgs scalar, $\chi_R \equiv [1,1,2,1]$, whose vacuum expectation value (VEV) breaks the 
left-right symmetry $G_{LR} \equiv SU(3)_c \times SU(2)_L \times SU(2)_R \times U(1)_X$ \cite{doub, doub1}. It is crucial to note that this field does not have any exclusive interaction with the SM fermions, and hence the $B-L$ quantum number is no longer uniquely determined as compared to the conventional LRSM. Therefore for $\chi_R$, we can choose $B-L=0$, and hence $\zeta=1$ in Eq. (\ref{1.7}). The left-right symmetry ensures that we have a second doublet Higgs scalar $\chi_L \equiv [1,2,1,1]$, with the same assignment of $B-L=0$ and $\zeta=1$. Interestingly, these assignments do not require any additional global symmetries, but will allow $B-L$ to remain as a global unbroken symmetry after the electroweak symmetry breaking.

\textit{A priori} we have two choices for the Higgs sector to break the electroweak symmetry. The first choice is that we keep the Higgs bi-doublet from the conventional model; after electroweak symmetry breaking 
it will then generate Dirac masses for all the fermions. Such a scenario is the subject of the discussion in Section \ref{app:0}. In this section we will be primarily interested in the alternative where there is no Higgs bi-doublet and left-handed Higgs doublet $\chi_L \equiv [1,2,1,1]$ breaks the electroweak symmetry. In such a scenario the quark masses and the charged lepton masses are generated through a seesaw mechanism that introduces new vector-like states \cite{fer-mass}. Interestingly, in this scenario the neutrino masses can be generated radiatively at the two loop level induced by $W_L-W_R$ mixing at the one loop level \cite{Babu:1988yq}. The field content of this model is summarised in Table I.
\begin{table}[t]
%[t!]
\begin{center}
\begin{tabular}{|c|c|c|c|c|c|c|}
\hline
Field     & $ SU(2)_L$ & $SU(2)_R$ & $B-L$ & $\zeta$ & $X=(B-L) +\zeta$ & $SU(3)_C$ \\
\hline
$q_L$        &  \bf{2}         & \bf{1}    & 1/3  & 0      & \bf{1/3}   & \bf{3}   \\
$q_R$       &  \bf{1}         & \bf{2}    & 1/3  & 0      & \bf{1/3}   &\bf{3}   \\
$\ell_L$     &  \bf{2}        & \bf{1}    & $-1$    & 0      & $\mathbf{-1}$    & \bf{1}   \\
$\ell_R$    &  \bf{1}         & \bf{2}    & $-1$    & 0      & $\mathbf{-1}$    & \bf{1}   \\
\hline
$U_{L,R}$ &  \bf{1}         & \bf{1}    & 1/3  & 1      & $\mathbf{4/3}$   & \bf{3}   \\
$D_{L,R}$ &  \bf{1}         & \bf{1}    & 1/3  &  $-1$    & $\mathbf{-2/3}$  & \bf{3}  \\
$E_{L,R}$ &  \bf{1}         & \bf{1}    & $-1$   & $-1$      & $\mathbf{-2}$    & \bf{1}   \\
\hline
 $\chi_L$   &  \bf{2}         & \bf{1}    & 0    & 1       & \bf{1}    & \bf{1}   \\
 $\chi_R$  &  \bf{1}        & \bf{2}    & 0   & 1        & \bf{1}    & \bf{1}   \\
 \hline
 $\rho$   &  \bf{1}         & \bf{1}    & 0    & 0       & \bf{0}    & \bf{1}   \\
\hline
\end{tabular}
\end{center}
\caption{Field content of the LRSM with an unbroken $B-L$ symmetry in the absence of a Higgs bi-doublet.}
{\label{tab:LR2}}
\end{table}

The necessity of the scalar field  $\rho$ in the model is justifiable from an examination of the relevant scalar potential~\cite{Mohapatra:1987nx}. In the absence of the Higgs bi-doublet the general scalar potential of this model can be written as 
\begin{eqnarray}{\label{eq:sp1}}
V=\,&& -\mu^2_\chi (\chi^{\dagger}_L \chi_L+\chi^{\dagger}_R \chi_R)
 + \lambda_1 [(\chi^{\dagger}_L \chi_L)^2+(\chi^{\dagger}_R \chi_R)^2]+\lambda_2 (\chi^{\dagger}_L \chi_L)
(\chi^{\dagger}_R \chi_R) -\mu^2_{\rho} \rho^2 +\lambda_\rho \rho^4\nonumber \\
&&~+
\mu_{\rho\chi} \rho(\chi^{\dagger}_L \chi_L -\chi^{\dagger}_R \chi_R)+\lambda_{\rho\chi} \rho^2(\chi^{\dagger}_L \chi_L+
\chi^{\dagger}_R \chi_R)\,.
\end{eqnarray}
Redefining $\lambda_1$ and $\lambda_2$ in terms of $\lambda_{+}=(\lambda_1 +\lambda_2/2)/2$ and  $\lambda_{-}=(\lambda_1 -\lambda_2/2)/2 $ and using the parametrisation $\langle\chi_L^{0}\rangle =r \sin \beta$, $\langle\chi_R^{0}\rangle =r \cos \beta$ and $\langle\rho\rangle =s $, we can recast the scalar potential in Eq. (\ref{eq:sp1}) as
\begin{eqnarray}{\label{eq:sp2}} 
V= -\mu^2_\chi r^2 +\lambda_{+} r^4 +\lambda_{-} r^4 \cos^2 2\beta - \mu^2_{\rho} s^2 +\lambda_\rho s^4 -\mu_{\rho\chi} s r^2 \cos 2\beta + \lambda_{\rho\chi}  s^2 r^2\,.
\end{eqnarray}
Minimising the scalar potential with respect to $r$, $\beta$ and $s$ we obtain
\begin{eqnarray}
-\mu^2_\chi+ 2\lambda_{+} r^2+2\lambda_{-} r^2 \cos^2 2\beta-\mu_{\rho\chi} s \cos 2\beta+ \lambda_{\rho\chi}  s^2&=0\,,
{\label{eq:sp3}} \\
\mu_{\rho\chi} r^2 s \sin 2\beta -2\lambda_{-} r^4 \cos 2\beta \sin 2\beta&=0\,,
{\label{eq:sp4}} \\
\lambda_{\rho\chi} r^2-\mu^2_{\rho}+2\lambda_\rho s^2-\mu_{\rho\chi} \cos 2\beta&=0\,.
{\label{eq:sp5}} 
\end{eqnarray}
From Eq. (\ref{eq:sp4}) it is evident that for $\mu_{\rho\chi}=0$, i.e. if the $\rho$ field is decoupled from the model then $\beta=\pi/4,\pi/2,\cdots$. Here $\beta=\pi/4$ corresponds to the unbroken parity symmetry case $\langle\chi_L^{0}\rangle =\langle\chi_R^{0}\rangle$ and $\beta=\pi/2$ corresponds to the case where $\langle\chi_L^{0}\rangle =0$, $\langle\chi_R^{0}\rangle\neq0$, leading to massless quarks and charged leptons. Therefore we conclude that it is crucial for the model to have a $\rho$ field with $\mu_{\rho\chi}\neq 0$ thus giving $\cos 2\beta=\mu_{\rho\chi} s/2\lambda_{-} r^2$, leading to a realistic mass spectrum for quarks and charged leptons of the model. Thus, we will consider the symmetry breaking pattern
\begin{eqnarray}\label{1.4.2}
&&SU(3)_c \times SU(2)_L \times SU(2)_R \times U(1)_{X} \times P~~[
\mathcal{G}_{LRP}] \nonumber \\
&\stackrel{\langle \rho \rangle}{\rightarrow}&SU(3)_c \times SU(2)_L \times SU(2)_R \times U(1)_X  ~~[
\mathcal{G}_{LR}] \nonumber \\
& \stackrel{\langle
\chi_R \rangle}{\rightarrow} &SU(3)_c \times
SU(2)_L \times U(1)_Y  ~~~~~~~~~~~~~~~~\,[ \mathcal{G}_{SM}] \nonumber \\
&\stackrel{ \langle \Phi \rangle}{\rightarrow} &SU(3)_c \times U(1)_Q  ~~~
~~~~~~~~~~~~~~~~~~~~~~~~~~~[ \mathcal{G}_{em}]\,.  \nonumber
\end{eqnarray}
In this scheme, the usual Dirac mass terms for the SM fermions are not allowed due to the absence of a Higgs bi-doublet scalar. However, under the presence of vector-like copies of quark and charged lepton gauge isosinglets, the charged fermion mass matrices can assume a seesaw structure. The relevant Yukawa interaction Lagrangian in this model is given by
\begin{eqnarray}
-\mathcal{L} = 
	&& ~h_{uL} \chi_L \overline{q}_L U_R 
	 + h_{uR} \chi_R  \overline{q}_R U_L +h_{dL} \tilde{\chi}_L \overline{q}_L D_R + h_{dR} \tilde{\chi}_R \overline{q}_R D_L + h_L \tilde{\chi}_L\overline{\ell}_L E_R 
	 + h_R \tilde{\chi}_R\overline{\ell}_R E_L \nonumber\\
	&&~+ m_U \overline{U}_L U_R + m_D \overline{D}_L D_R + m_E \overline{E}_L E_R + \text{h.c.}\, ,
\label{2.1}
\end{eqnarray}
where we suppress the flavour and colour indices on the fields and couplings for brevity. $\tilde{\chi}_{L,R}$ denotes $\tau_2 \chi_{L,R}^\ast$, where $\tau_2$ is the usual second Pauli matrix. Note that, in general, if the parity symmetry is broken by the VEV of a singlet scalar at some high scale as compared to the left-right symmetry breaking scale then the Yukawa couplings corresponding to the right-type and left-type Yukawa terms may run differently under the renormalisation group below the parity breaking scale. This approach where the left-right parity symmetry and $SU(2)_R$ breaking scales are decoupled from each other was first proposed in \cite{Chang:1983fu}. Therefore, while writing the Yukawa terms above we distinguish the left- and right-handed couplings explicitly with the subscripts $L$ and $R$.

After electroweak symmetry breaking we can write the mass matrices for the quarks as \cite{Deppisch:2016scs,Dev:2015vjd,Dasgupta:2015pbr,Deppisch:2017vne,Patra:2017gak}
\begin{eqnarray}
\label{2.3}
	M_{uU}    = \pmatrix{ 0 & h_{uL} u_L \cr h_{uR}^\dagger u_R & m_U }, \, \, \, \,
	M_{dD}    = \pmatrix{ 0 & h_{dL} u_L \cr h_{dR}^\dagger u_R & m_D }\, .
\end{eqnarray}
where $\langle \chi_{L,R}\rangle =u_{L,R}$. Up to leading order in $h_{uL} u_L$, the SM and heavy vector partner up-quark masses are given by
\begin{eqnarray}
\label{eq001}
	m_u \approx h_{uL} h_{uR}  \frac{u_L u_R}{m_U}\,, \quad
	\hat{m}_U \approx \sqrt{m_U^2 + (h_{uR} u_R)^2}\,,
\end{eqnarray}
\textit{A priori}, the up type quark mass matrices can be diagonalised via left and right unitary transformations giving rise to the usual Cabibbo-Kobayashi-Maskawa (CKM) matrix and its right handed analog, in the basis where down-type quark mass matrix is already diagonal. Simplified expressions for the mixing angles $\theta^{L,R}_U$ can be found in the limit where the Yukawa couplings are assumed to be real and therefore the diagonalising unitary matrices are simplified to orthogonal matrices ${\cal{O}}^{L,R}$. In this case the mixing angles $\theta^{L,R}_U$ are given by
\begin{eqnarray}
\label{eqoo2}
	\tan(2\theta^{L,R}_U) \approx 2 h_{uL,uR} \frac{u_{L,R} m_U}{m_U^2 \pm (h_{uR} u_R)^2}\,.
\end{eqnarray}
The down-quark masses and mixing are obtained in an analogous manner. Note that in writing the above equations we have dropped the flavour indices of the Yukawa couplings $h_{uL,uR}$ which determine the observed quark and charged lepton mixings. The hierarchy of the quark masses can be explained by assuming either a hierarchical structure of the Yukawa couplings or a hierarchical structure of more than one generation of the vector-like quark masses.

Similarly, the charged lepton masses are generated through a Dirac seesaw mechanism. However, we explicitly assume more than one generation of vector-like charged lepton and work in a basis where the vector-like charged lepton masses are diagonal. In such a basis the SM charged lepton masses are given by 
\begin{equation}{\label{lepton:mass1}}
    m_{l_{ij}} = u_L u_R h_{L_{ik}} M^{-1}_{E_{k}} h^{\dagger}_{R_{kj}}\,.
\end{equation}
The charged lepton mass matrix given in Eq. (\ref{lepton:mass1}) can be diagonalised by the bi-unitary transformation
\begin{eqnarray}{\label{CLMNS}}
m^{\text{diag}}_{l_{\alpha}}=U_{L_{\alpha i}}^{l\dagger}m_{l_{ij}} U^l_{R_{j \alpha}}\,,
\end{eqnarray}
where $l^m_{L(R)}=U_{L(R)}l^f_{L(R)}$ with the superscripts $m$ and $f$ correspond to the mass and flavour bases, respectively.
\begin{figure}[b]
\includegraphics[width=3 in]{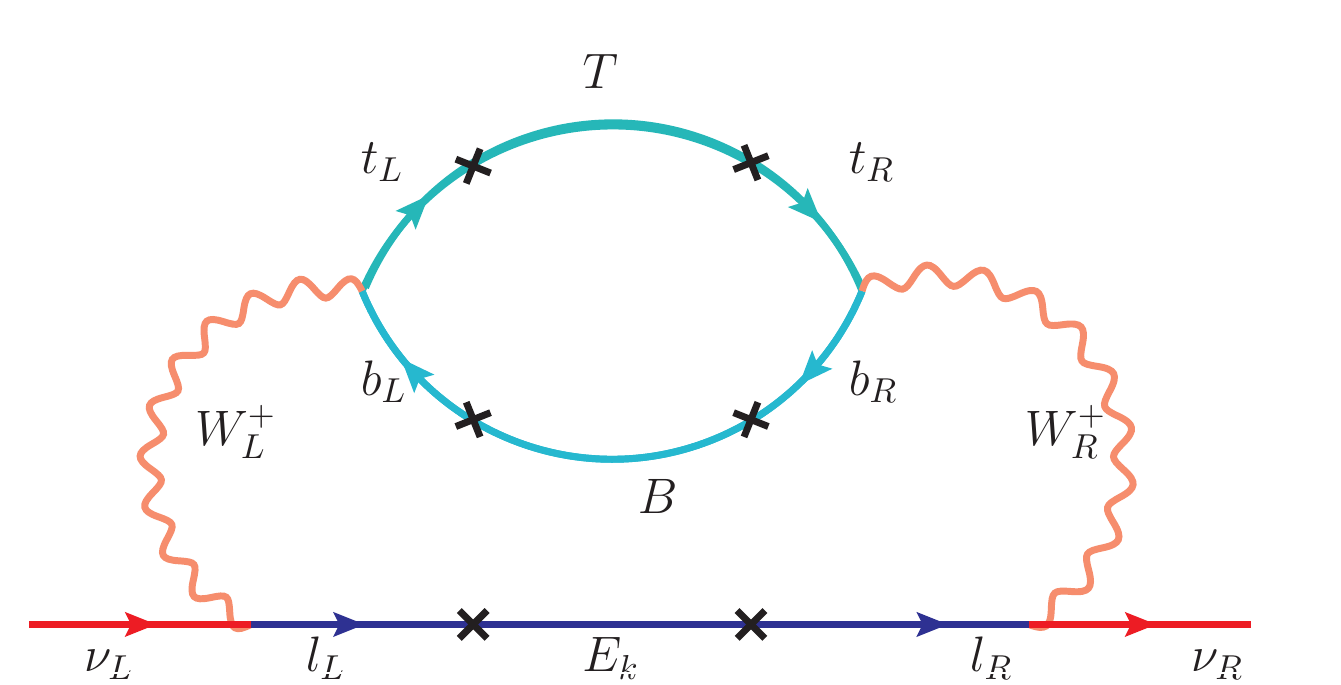}
\caption{Two-loop radiative diagram generating Dirac neutrino masses.}
\label{fig:numass}
\end{figure}
Light Dirac neutrino masses are generated through a two loop contribution employing the mixing of $W_L$ and $W_R$ occurring at a one loop level \cite{Babu:1988yq}. The relevant Feynman diagram is shown in Fig.~1. The computation of this diagram leads to the following neutrino mass term
\begin{eqnarray}{\label{eq:numass}}
\mathcal{L} \, =\, && - \frac{g_L^2 g_R^2}{2} \frac{m_B m_T}{m_{W_L}^2 m_{W_R}^2} h_{tL}h_{tR}^{\dagger}  h_{bL}h_{bR}^{\dagger} u_L^3 u_R^3 \, \bar{\nu}_{L_i} h_{L_{ik}} \mathcal{I}'_{k} h^{\dagger}_{R_{kj}}  {\nu_{R_j}}\,,
\end{eqnarray}
where $\mathcal{I}'_k= m_{E_k} \mathcal{I}_k$ corresponds to a diagonal matrix, with
\begin{eqnarray}{\label{eq:integral}}
\mathcal{I}_k= \int \frac{d^{4}k}{\left(2\pi\right)^{4}}\,
\int \frac{d^{4}p}{\left(2\pi\right)^{4}}\,
\frac{3 m_{W_L}^2 m_{W_R}^2 +(p^2-m_{W_L}^2)(p^2-m_{W_R}^2)}{p^{2} (p^{2}-m_{E_{k}}^2)  (p^2-m_{W_L}^2)(p^2-m_{W_R}^2) k^2 (k^2-m_B^2) (p+k)^2 [(p+k)^2-m_{T}^2]}\, .\nonumber\\
\end{eqnarray}
Here $p$ and $k$ denote the momenta of the $W_L$ and $b$ in the loops, respectively. Note that to simplify the analysis we have made the assumption that the top and the bottom quarks contribute dominantly in the one loop diagram inducing the mixing between $W_L$ and $W_R$, and consequently in writing Eq. (\ref{eq:numass}) we treat the corresponding Yukawa couplings $h_t$ and $h_b$ as numbers instead of matrices in the presence of more than one generation of heavy vector-like quarks. On the other hand, $h_{L,R}$ are in general $3\times 3$ matrices which play a crucial role in understanding the neutrino masses and mixings. We would like to point out that for a scenario with a single generation of vector-like charged leptons or more than one generation of vector-like charged leptons with degenerate masses in the integral given in Eq. (\ref{eq:integral}) the neutrino mass matrix turns out to be directly proportional to the charged lepton mass matrix and consequently, the PMNS matrix turns out to be diagonal which is ruled out by the current neutrino oscillation data\footnote{In such scenarios the situation can be remedied by extending the field content of the model to also include heavy vector-like neutrinos to realise a Dirac seesaw scenario or by extending the Higgs sector to realise a one loop radiative mechanism for generating the neutrino masses.}. 
However, we would like to emphasise that the above argument is no longer true in the case where more than one generation of vector-like charged leptons with a hierarchical mass spectrum is considered. In Section \ref{sec:para}, we shall focus on this scenario and show that it is indeed possible to accommodate non-trivial mixings in the PMNS mixing matrix using only the two loop radiative contribution if more than one generation of heavy vector-like charged lepton is present. 

In Appendices \ref{app:A} and \ref{app:B} we sketch two alternative methods of evaluating the two loop integral given in Eq. (\ref{eq:integral}). Note that even though such integrals have been evaluated in the literature for one heavy vector-like charged lepton state under some simplifying assumptions \cite{Babu:1988yq}, it is crucial to evaluate them more generally to understand the dependence of the integral on the vector-like quark masses, which generate a non-trivial mixing for the neutrinos in addition to nonzero masses. Following the approach outlined in Appendix \ref{app:A}, the final neutrino masses are given by
\begin{eqnarray}{\label{eq:numass2}}
m_{\nu_{ij}} \, =\, &&  \frac{g_L^2 g_R^2}{2} \frac{m_b m_t}{m_{W_L}^2 m_{W_R}^2}  u_L u_R \, h_{L_{ik}} \mathcal{J}_{k} h^{\dagger}_{R_{kj}}\,,
\end{eqnarray}
where
\begin{eqnarray}{\label{eq:integral2}}
\mathcal{J}_k= \frac{m_{E_k}}{(16 \pi^2)^2}\int^{\infty}_{0} \,dr \,\frac{\alpha_k}{r+\alpha_k}\,
\int ^{1}_{0}\, dx\, \ln\left[ 
\frac{x(1-x)r+(1-x)}{x(1-x)r+(1-x)+x\beta_k} \,\frac{(1-x)r+\beta_k }{(1-x)r}
\right]\,,
\end{eqnarray}
with $\alpha_k=m_B^2/m_{E_k}^2$ and $\beta_k=m_T^2/m_{E_k}^2$. The neutrino mass matrix given in Eq. (\ref{eq:numass2}) can be diagonalised by the bi-unitary transformation
\begin{eqnarray}{\label{NMNS}}
m^{\text{diag}}_{\nu_{\alpha}}=U_{L_{\alpha i}}^{\nu\dagger}m_{\nu_{ij}} U_{R_{j \alpha}}^{\nu}\,,
\end{eqnarray}
where $U^{\nu}_{L}$ and $U^{\nu}_{R}$ are the left- and right-handed unitary matrices corresponding to the bi-unitary transformation diagonalising the neutrino mass matrix. 
%%%%%%%%%%%%%%%%%%%%%%%%%%%%%%%%%%%%%%%%%%%%%%%%%%%%%%%%%%%%%%%
\section{A left-right symmetric parametrisation of the radiative neutrino masses and mixing}{\label{sec:para}}
To analyse the two loop radiative neutrino masses and mixings phenomenologically, it is convenient to parametrise the charged lepton and neutrino masses. From Eqs. (\ref{lepton:mass1}) and (\ref{CLMNS}) the diagonal charged lepton matrix is given by
\begin{eqnarray}{\label{ld}}
{\bf{m^{\text{diag}}_l}}= \bf{U}_{L}^{l\dagger}  h_{L} \hat{M}^{-1}_{E} h^{\dagger}_{R}  U^l_{R}\,,
\end{eqnarray}
 where the matrices have been made bold to distinguish them from numbers and ${\bf{\hat{M}^{-1}_{E}}} =u_L u_R {\bf{m_E^{-1}}}$ is a diagonal matrix. Similarly, from Eqs. (\ref{eq:numass2}) and (\ref{NMNS}) the diagonal neutrino mass matrix is given by
\begin{eqnarray}{\label{nud}}
{\bf {m^{\text{diag}}_{\nu}}}={\bf U_{L}^{\nu\dagger} h_{L} M_{E_\nu} h^{\dagger}_{R} U_{R}^{\nu}}\,,
\end{eqnarray}
where
\begin{eqnarray}{\label{nud2}}
{\bf{M_{E_\nu}}}= \frac{g_L^2 g_R^2}{2} \frac{m_b m_t}{m_{W_L}^2 m_{W_R}^2}  u_L u_R \,{\bf J }
\end{eqnarray}
is a diagonal matrix with ${\bf{J}}$ being the diagonal matrix corresponding to the integral Eq. (\ref{eq:integral2}). If  ${\bf{J}}$ is not proportional to ${\bf{m_E^{-1}}}$ then one can have a non-trivial PMNS mixing matrix ${\bf{U_{L}=U_{L}^{l\dagger} U_{L}^{\nu}}}$ by solving Eqs. (\ref{ld}) and (\ref{nud}) simultaneously, in order for $h_{L}$ and $h_{R}$ to fit the neutrino oscillation data. A comprehensive numerical analysis of the $3\times 3$ general left-right asymmetric mixing case is highly non-trivial and involves a large number of parameters. This is beyond the scope of the current work and will be addressed in a future work. Here we will focus on a particularly interesting limiting case where the left- and right-handed unitary rotation matrices and the Yukawa couplings are identical i.e. ${\bf{U_{L}^{l,\nu}\equiv U_{R}^{l,\nu}} \equiv U^{l,\nu}}$ and ${\bf{ h_{L}\equiv h_{R} \equiv h }}$. This helps us to construct a new parametrisation \`{a} la Casas-Ibarra~\cite{Casas:2001sr} which immensely simplifies the underlying numerical analysis of simultaneously solving Eqs. (\ref{ld}) and (\ref{nud}). Even though such a simplifying assumption need not be true in general, it enables us to explore the qualitative dependence of the PMNS mixing angle on different model parameters by using a phenomenological approach. As noted before, for a diagonal ${ \bf m_E} $ and ${\bf J}$, ${\bf{\hat{M}^{-1}_{E} }}$ and ${\bf{M_{E_\nu}}}$ are diagonal matrices in generation space, which allows us to write the identities
\begin{eqnarray}
\bf({m^\text{diag}_l}^{-1/2} U^{l\dagger}\; h\; \hat{M}^{-1/2}_{E})\,
(\hat{M}^{-1/2}_{E} \; h^{\dagger}\;
U^{l}\;{m^\text{diag}_l}^{-1/2})\,&=&\,
\mathbb{I}\,=\,{\cal R}_l\,{\cal R}_l^{\dagger}\,, {\label{eq:cip1}}\\
\bf({m^\text{diag}_\nu}^{-1/2} U^{\nu\dagger}\; h\; M_{E_\nu}^{1/2})\,
(M_{E_\nu}^{1/2}\; h^{\dagger}\;
U^{\nu}\;{m^\text{diag}_\nu}^{-1/2})\,&=&\,
\mathbb{I}\,=\,{\cal R}_\nu\,{\cal R}_\nu^{\dagger}\,, 
{\label{eq:cip2}}
\end{eqnarray}
where ${\cal R}_{l,\nu}$ are arbitrary unitary matrices (${\cal R}^{\dagger}{\cal R}=\mathbb{I}$). Working in a basis where the charged lepton masses are diagonal, i.e. ${\bf U^{l}=\mathbb{I}}$ and ${\bf U^{\nu} \equiv U}$, the PMNS mixing matrix, one can solve Eq. (\ref{eq:cip1}) for the Yukawa matrix {\bf h} up to an arbitrary unitary matrix ${\cal R}_{l}$
\begin{eqnarray}{\label{eq:cip3}}
{\bf h }\,=\, {\bf {m^\text{diag}_l}^{1/2}} \,{\cal R}_l\, {\bf \hat{M}^{1/2}_{E}}\,,
\end{eqnarray}
which can then be substituted into Eq.~(\ref{eq:cip2}) to solve for the PMNS mixing matrix up to an arbitrary unitary matrix ${\cal R}_{\nu}$
\begin{eqnarray}{\label{eq:cip4}}
\bf U\,=\,(h^{\dagger})^{-1}\, M_{E_\nu}^{-1/2} \,{\cal R}_\nu^{\dagger} \, {m^\text{diag}_\nu}^{1/2}\,. 
\end{eqnarray}
%
%%%%%%%%%%%%%%%%%%%%%%%%%%%%%%%%%%%%%%%%%%%%%%%%%%%%%%%%%%%%%%%
In order to understand the dependence of the PMNS mixing angle on different model parameters (in particular, the left-right symmetry breaking scale and mass scale of the heavy vector-like fermions) and the arbitrary unitary rotation matrices qualitatively, we explore the discussed parametrisation to solve Eq. (\ref{eq:cip3}) and (\ref{eq:cip4}) simultaneously for a $2\times2$ case. Furthermore we restrict ourselves to the case where all the Yukawa matrices and rotation matrices are real. With these simplifying assumptions, the arbitrary rotation matrices ${\cal R}_{l,\nu}$ and the PMNS matrix $U$ can now be parametrised in terms of one rotation angle each
\begin{eqnarray}{\label{eq:na1}}
{\bf U}= \pmatrix{\cos \theta \hfill & \sin \theta \cr -\sin \theta & \cos \theta \hfill }, \,
{\mathcal{R}_l}= \pmatrix{\cos \theta_l \hfill & \sin \theta_l \cr -\xi \sin \theta_l & \xi \cos \theta_l \hfill }, \,
{\mathcal{R}_\nu}= \pmatrix{\cos \theta_\nu \hfill &  \sin \theta_\nu \cr -\xi \sin \theta_\nu & \xi \cos \theta_\nu \hfill },
\end{eqnarray}
where $\xi=\pm 1$, and $\theta$ corresponds to the usual PMNS maximal angle $\theta_{23}$. Among the other free parameters we set $g_R=g_L$, $m_T =1.5$~TeV to be consistent with the current search limits from \cite{Sirunyan:2018qau,Aaboud:2018ifs,Aaboud:2018pii} and the lightest vector-like charged lepton mass $m_{E_1}=1$~TeV to be consistent with the current search limits from \cite{CMS:2018cgi}, as benchmark points. For the $2 \times 2$ matrix ${\bf {m^\text{diag}_l}}$ we choose the diagonal entries to be muon and tau masses. Further, the $2\times2$ approximation makes use of the hierarchy of mass squared splittings -- the diagonal entries of $\mathbf{m}^\text{diag}_{\nu}$ are set by the splittings, while the $2\times 2$ mixing angle $\theta$ corresponds approximately to the  $3\times 3 $ atmospheric mixing angle $\theta_{23}$. We use the best-fit values for the atmospheric and solar neutrino mass squared differences from the global oscillation analysis \cite{deSalas:2017kay}. For ease of reference, the relevant global analysis parameters are summarised in Table \ref{tab:os}. For these benchmark choices, we solve Eqs. (\ref{eq:cip3}) and (\ref{eq:cip4}) simultaneously to obtain simultaneous solutions for four parameters $\theta$, $u_R$, $\theta_l$ and $\theta_l$ as a function of the mass difference between two generations of vector-like charged lepton masses $m_{E_2}-m_{E_1}\equiv \Delta m_E$ for different benchmark values of $m_B$ . 
\begin{table}[]
	\begin{tabular}{|l|l|l|l|}
		\hline
		~~~~~~~Parameter & ~~Best Fit $\pm ~1 \sigma$~~ & ~~~~~~~~~~~Parameter &  ~~Best Fit $\pm ~1 \sigma$~~ \\ \hline
		~~~~~~$\sin ^{2} \theta_{12} / 10^{-1}$ & ~~~~~$3.20^{+0.20}_{-0.16}$ & ~~~~~~~~~~$\delta_{\mathrm{CP}} / \pi~(\mathrm{NO})$ & ~~~~~$1.21^{+0.21}_{-0.15}$ \\ \cline{1-2}
		~~$\sin ^{2} \theta_{23} / 10^{-1}~(\mathrm{NO})$~~&  ~~~~~$5.47^{+0.20}_{-0.30}$  & ~~~~~~~~~~$\delta_{\mathrm{CP}} / \pi~(\mathrm{IO})$ & ~~~~~$1.56^{+0.13}_{-0.15}$  \\ \cline{3-4}
		~~$\sin ^{2} \theta_{23} / 10^{-1}~(\mathrm{IO})$&   ~~~~~$5.51^{+0.18}_{-0.30}$ & ~~~~~~~$\Delta m_{21}^2 ~[10^{-5} ~\mathrm{eV}^2]$ & ~~~~~$7.55^{+0.20}_{-0.16}$   \\ \hline
		~~$\sin ^{2} \theta_{13} / 10^{-2}~(\mathrm{NO})$& ~~~~$2.160^{+0.083}_{-0.069}$ & ~~$\left|\Delta m_{31}^{2}\right|\left[10^{-3} ~\mathrm{eV}^{2}\right]~(\mathrm{NO})$~~ & ~~~~$2.50\pm0 .03$  \\ 
		~~$\sin ^{2} \theta_{13} / 10^{-2}~(\mathrm{IO})$& ~~~~$2.220^{+0.074}_{-0.076}$ & ~~~$\left|\Delta m_{31}^{2}\right|\left[10^{-3} ~\mathrm{eV}^{2}\right](\mathrm{IO})$~~ &  ~~~~~$2.42^{+0.03}_{-0.04}$ \\ \hline
	\end{tabular}
\caption{Current global best-fit values for the neutrino oscillation parameters, taken from \cite{deSalas:2017kay}.}
{\label{tab:os}}
\end{table}
\begin{figure}[t]{\label{fig:NH}}
\includegraphics[width=3.2 in]{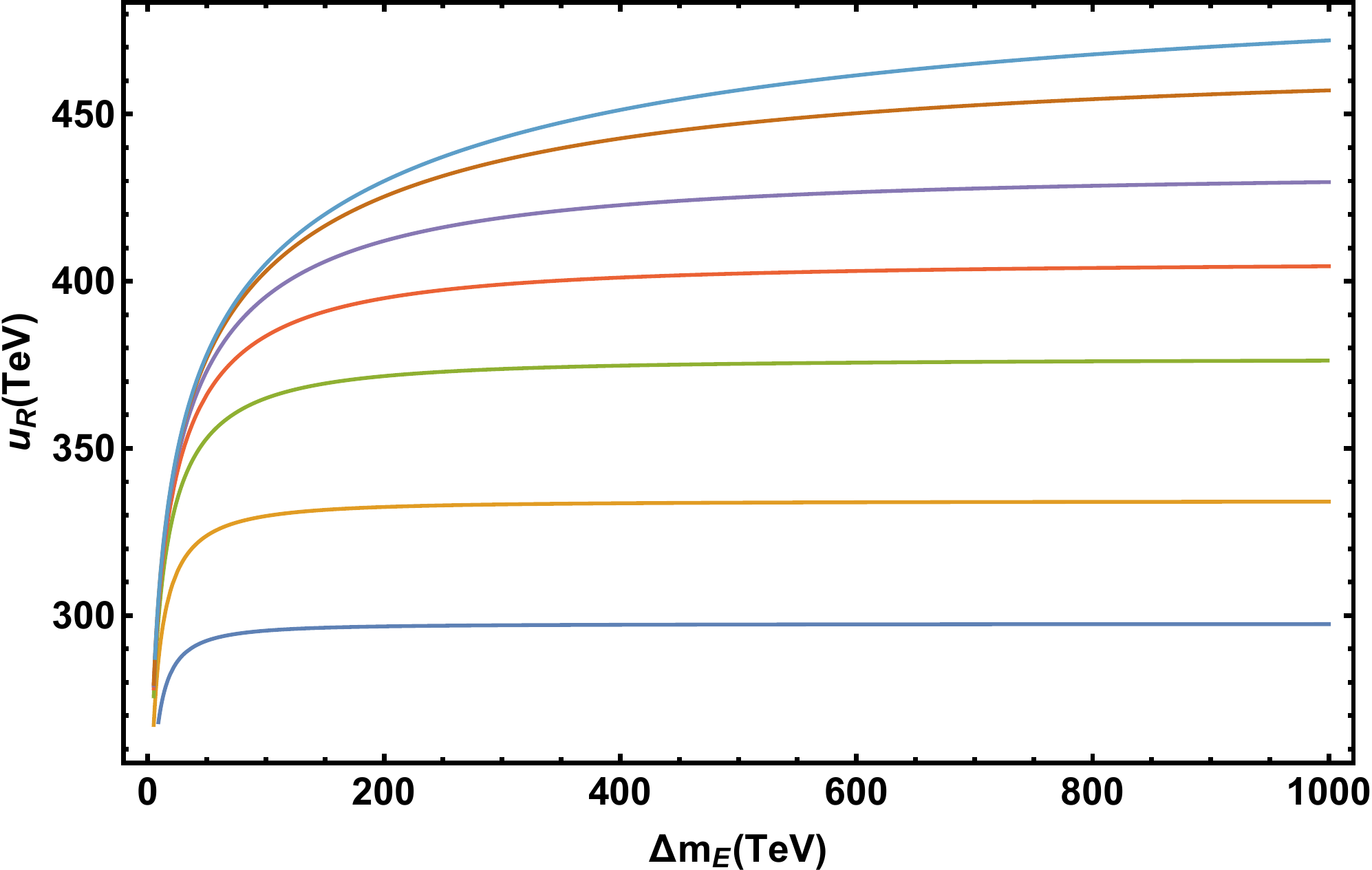}
\includegraphics[width=3.2 in]{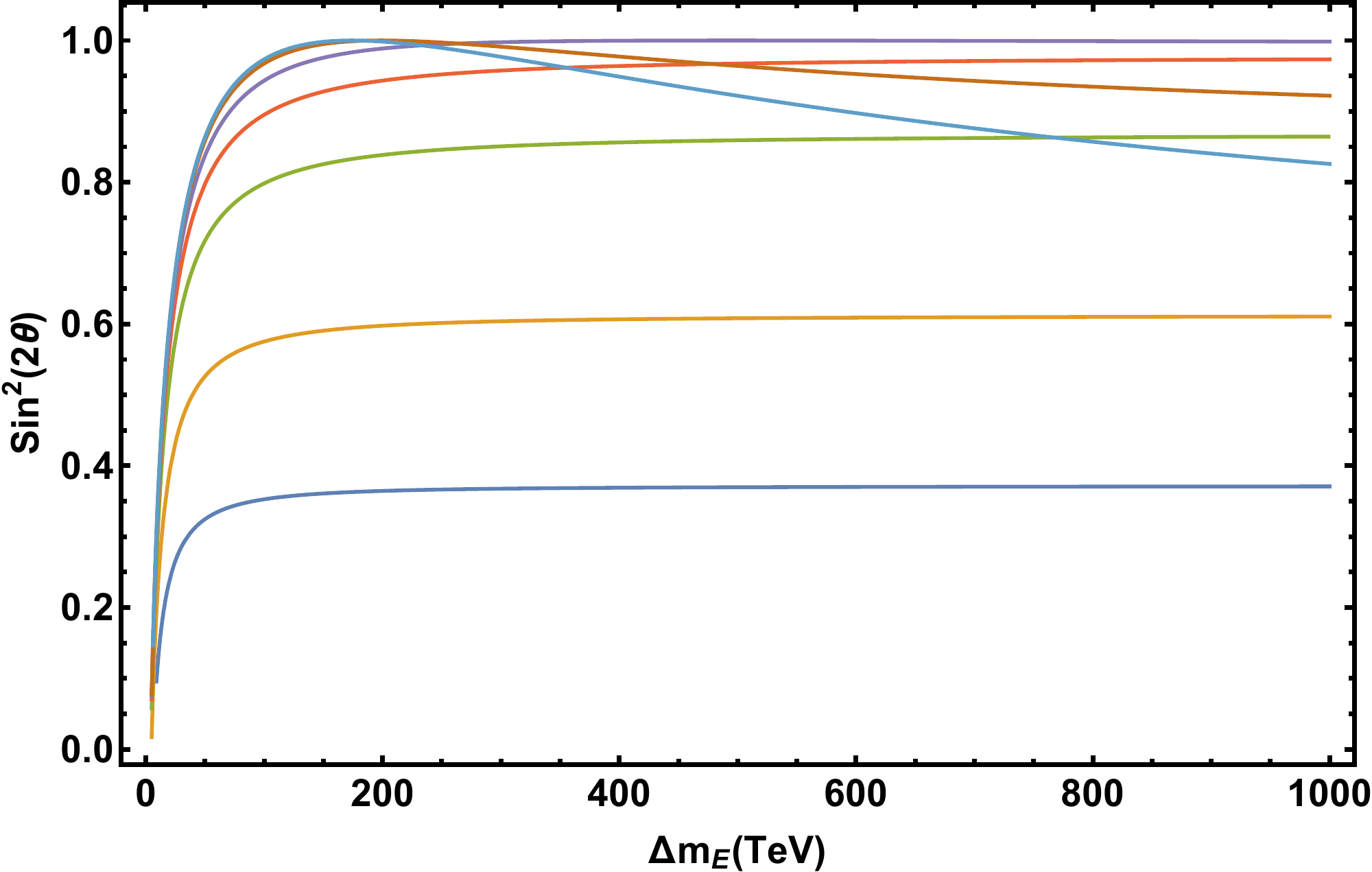}
\includegraphics[width=3.2 in]{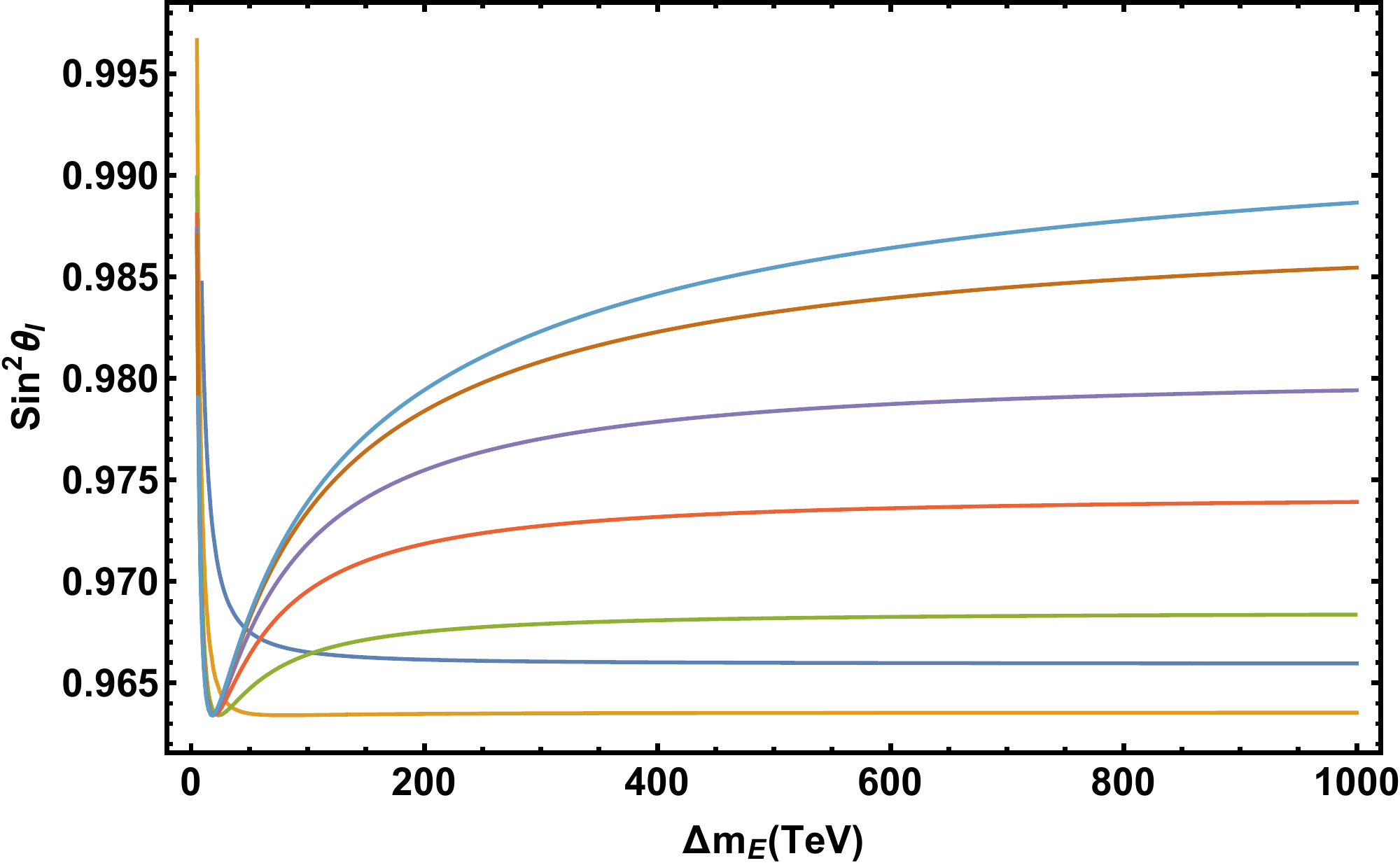}
\includegraphics[width=3.2 in]{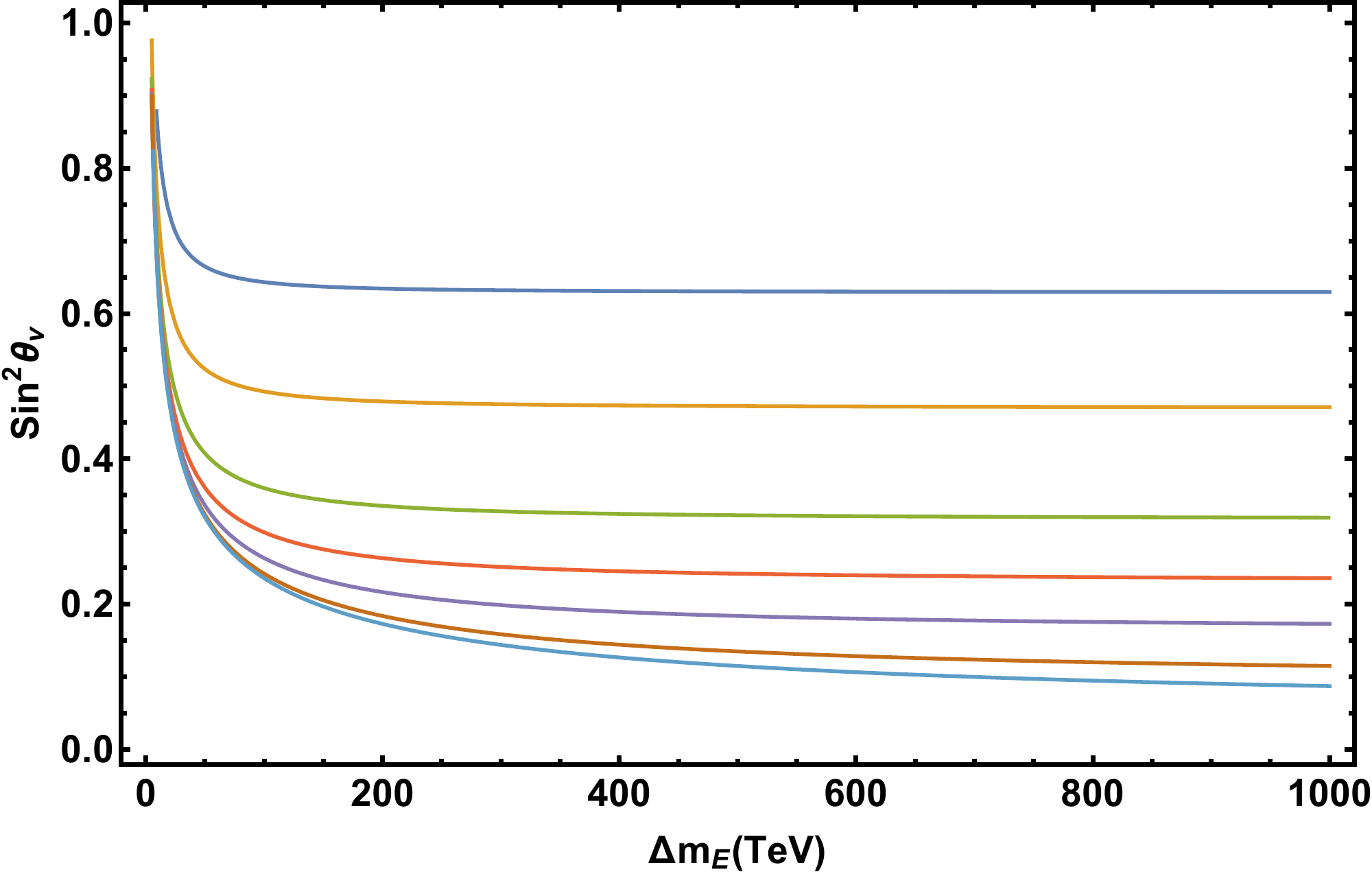}
\includegraphics[width=3.2 in]{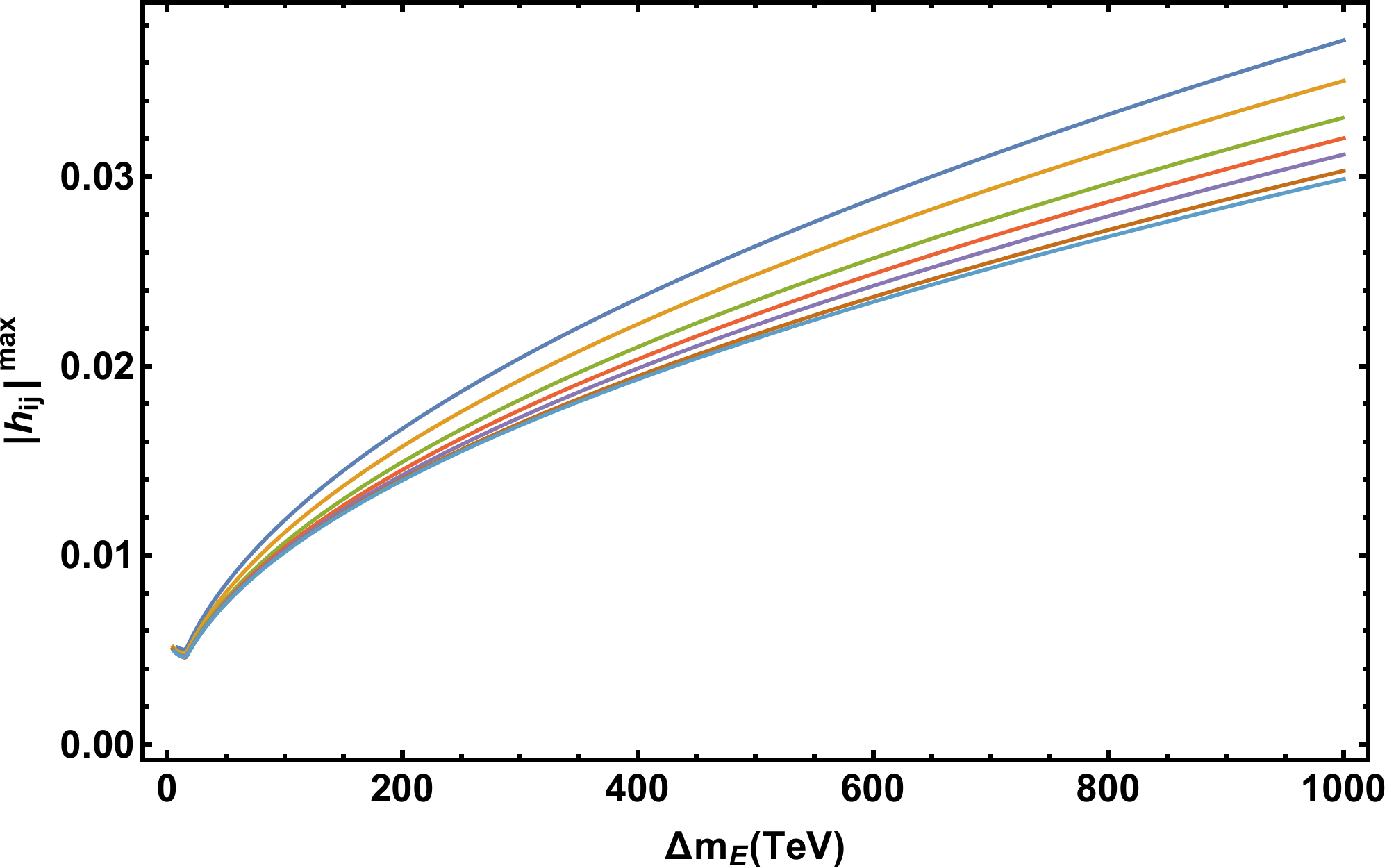}
\includegraphics[width=1. in]{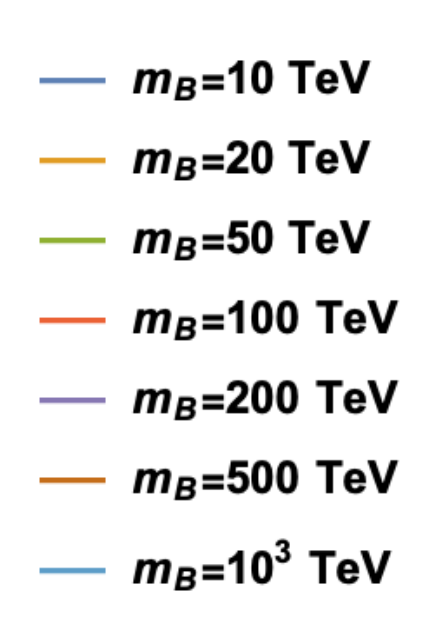}
\caption{Numerical solutions for $2\times 2$ mixing simultaneously fitting Eqs. (\ref{eq:cip3}) and (\ref{eq:cip4}) for the case of normal ordering of neutrino masses: (top left) $SU(2)_R$ breaking scale $u_R$,  (top right) $2\times 2$ PMNS mixing angle, (middle left) the arbitrary rotation matrix angles in ${\mathcal{R}_l}$, (middle right) the arbitrary rotation matrix angles in ${\mathcal{R}_\nu}$, (bottom) the maximal element of the Yukawa matrix ${\bf h}$, as a function of the mass difference between two generations of vector-like charged lepton masses $\Delta m_E$ for different benchmark values of $m_B$. See text for the benchmark values of the other relevant parameters.}
\end{figure}
\begin{figure}[t]{\label{fig:IH}}
\includegraphics[width=3.2 in]{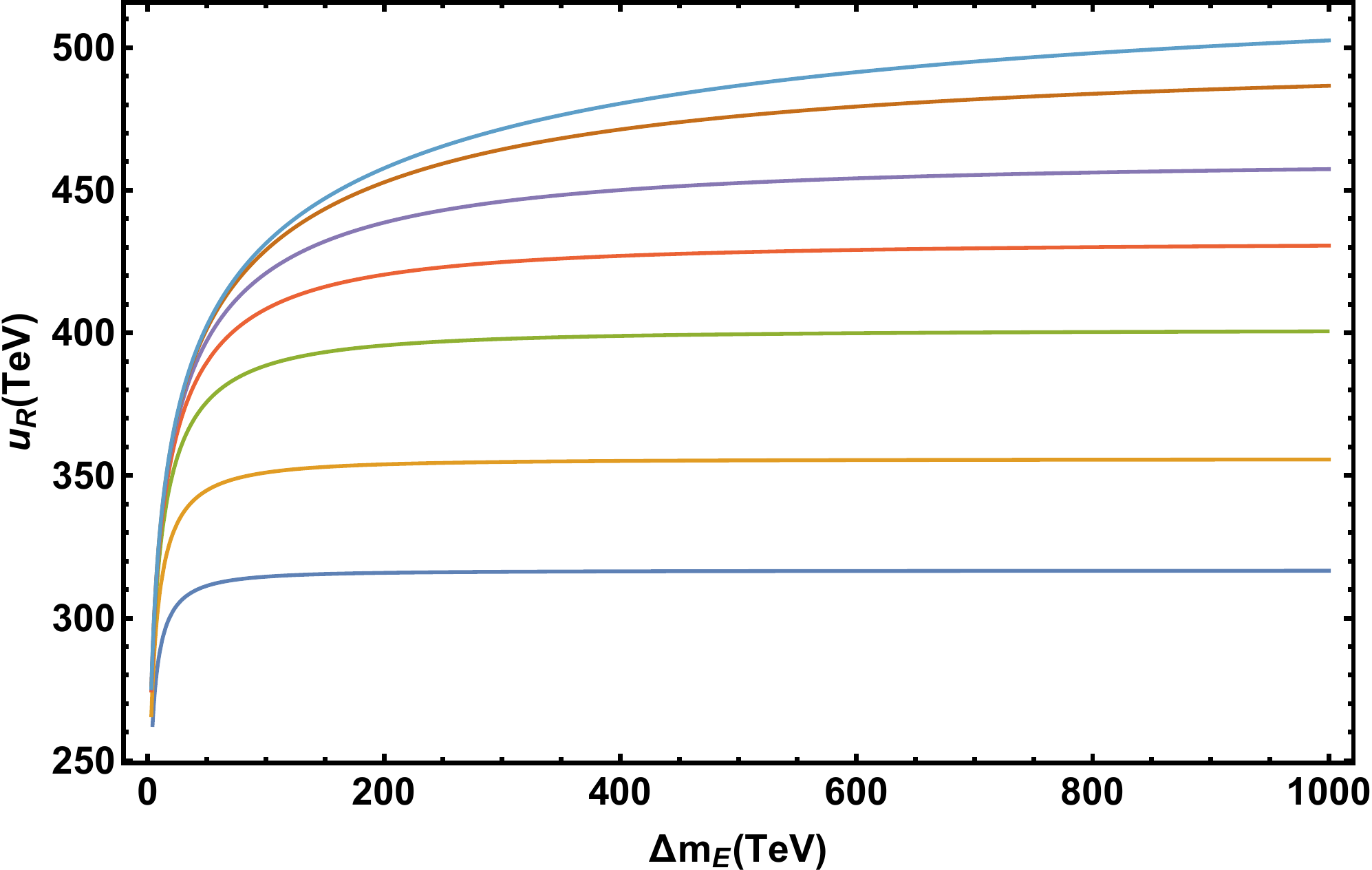}
\includegraphics[width=3.2 in]{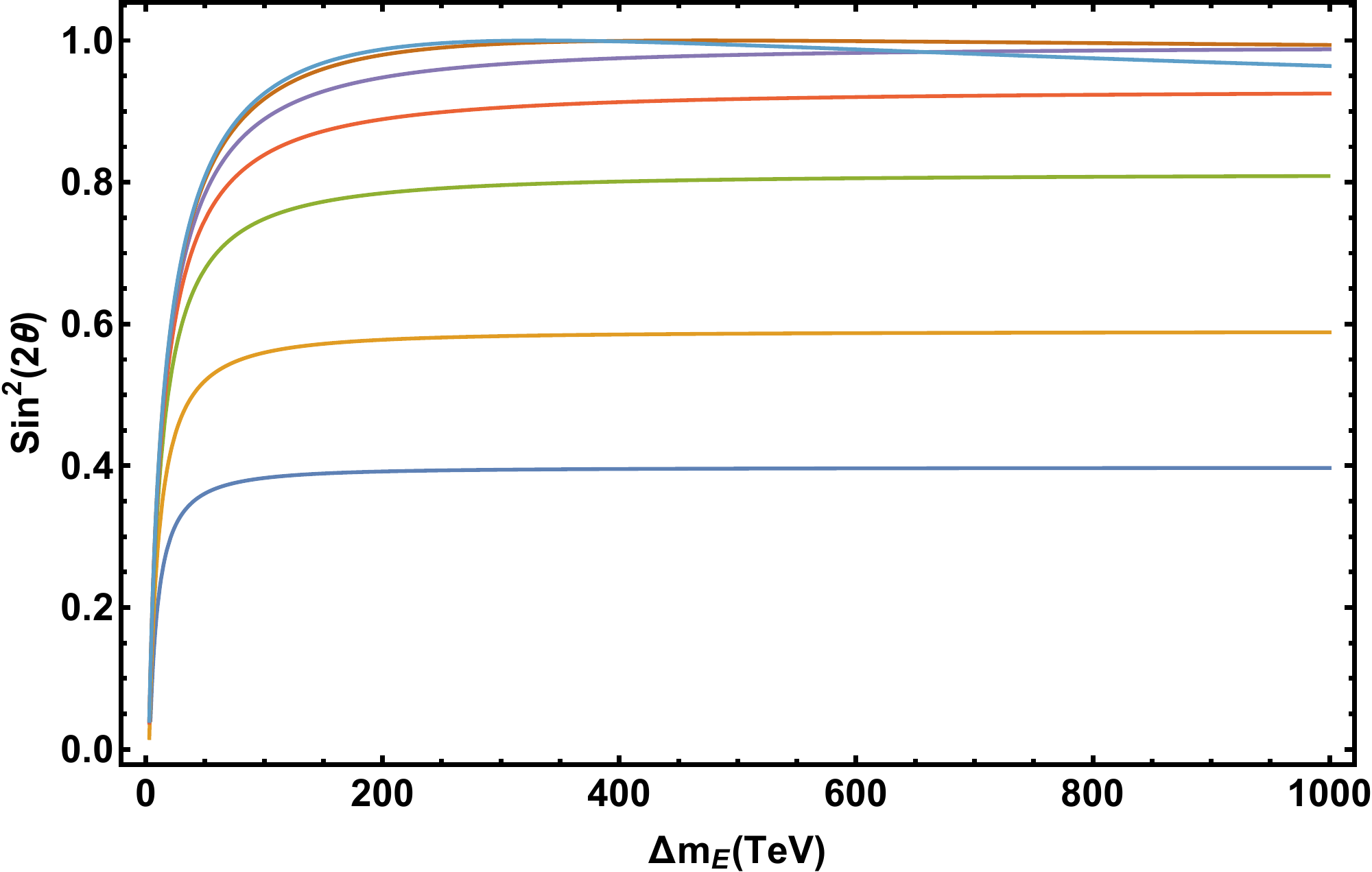}
\includegraphics[width=3.2 in]{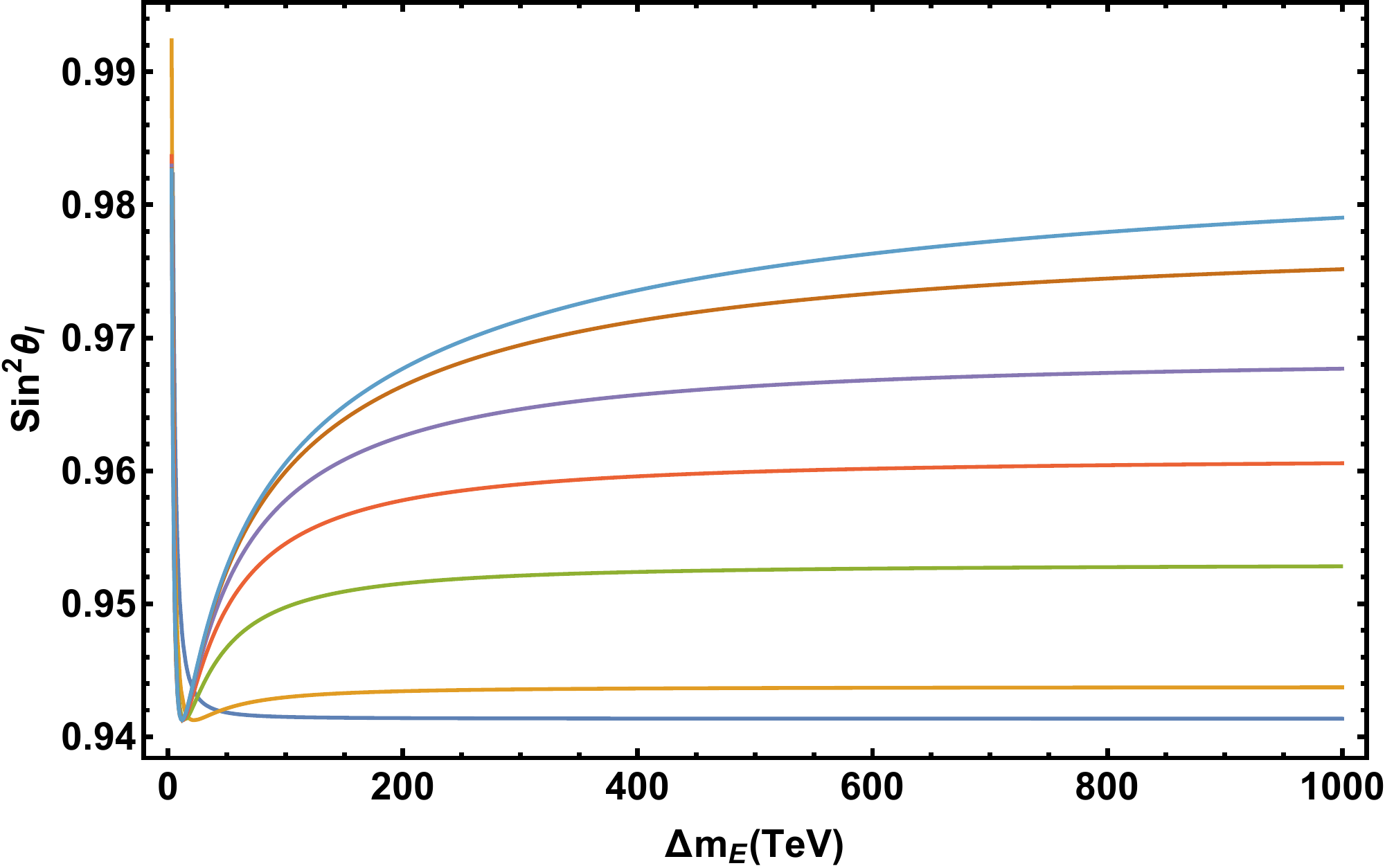}
\includegraphics[width=3.2 in]{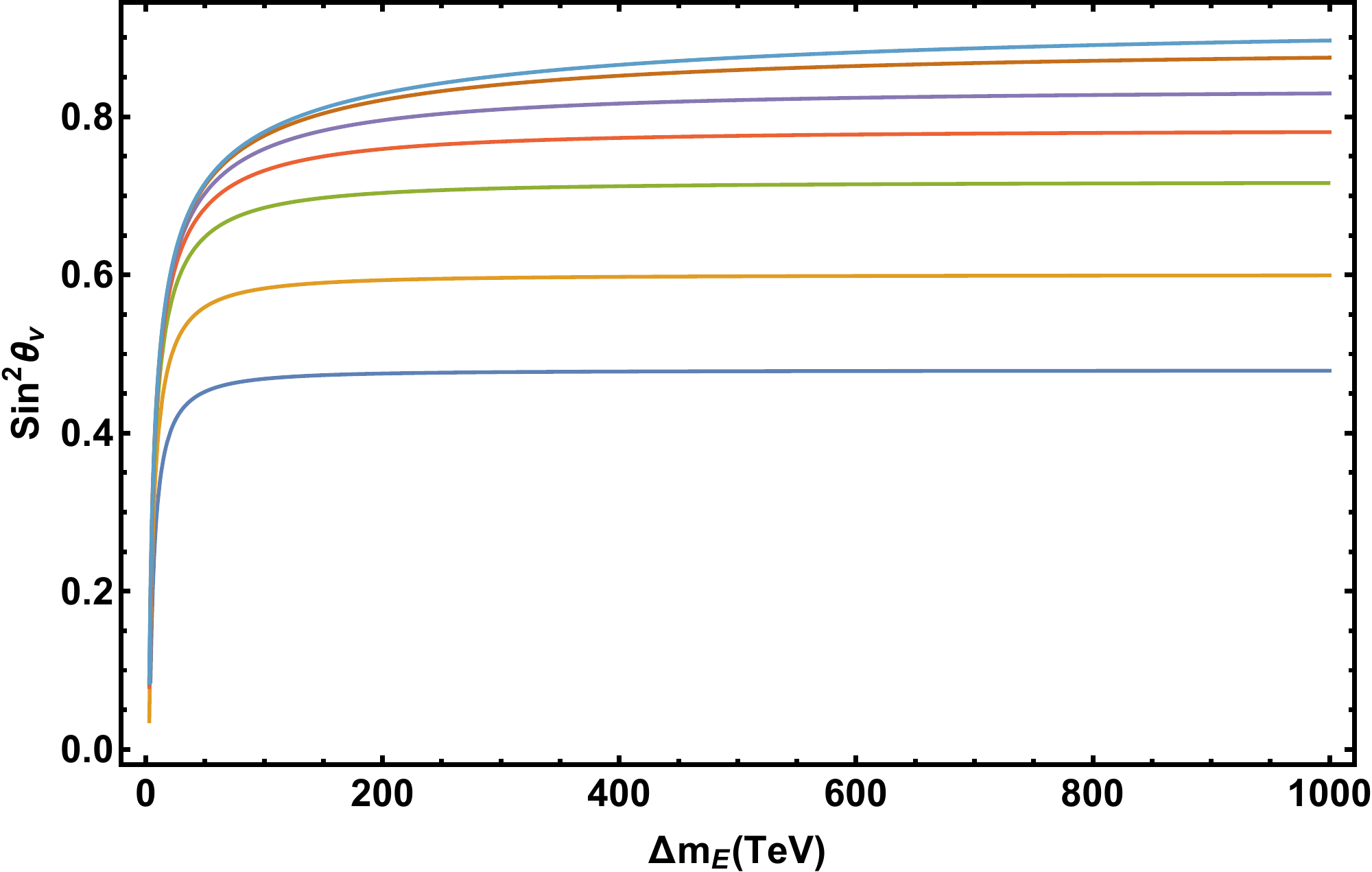}
\includegraphics[width=3.2 in]{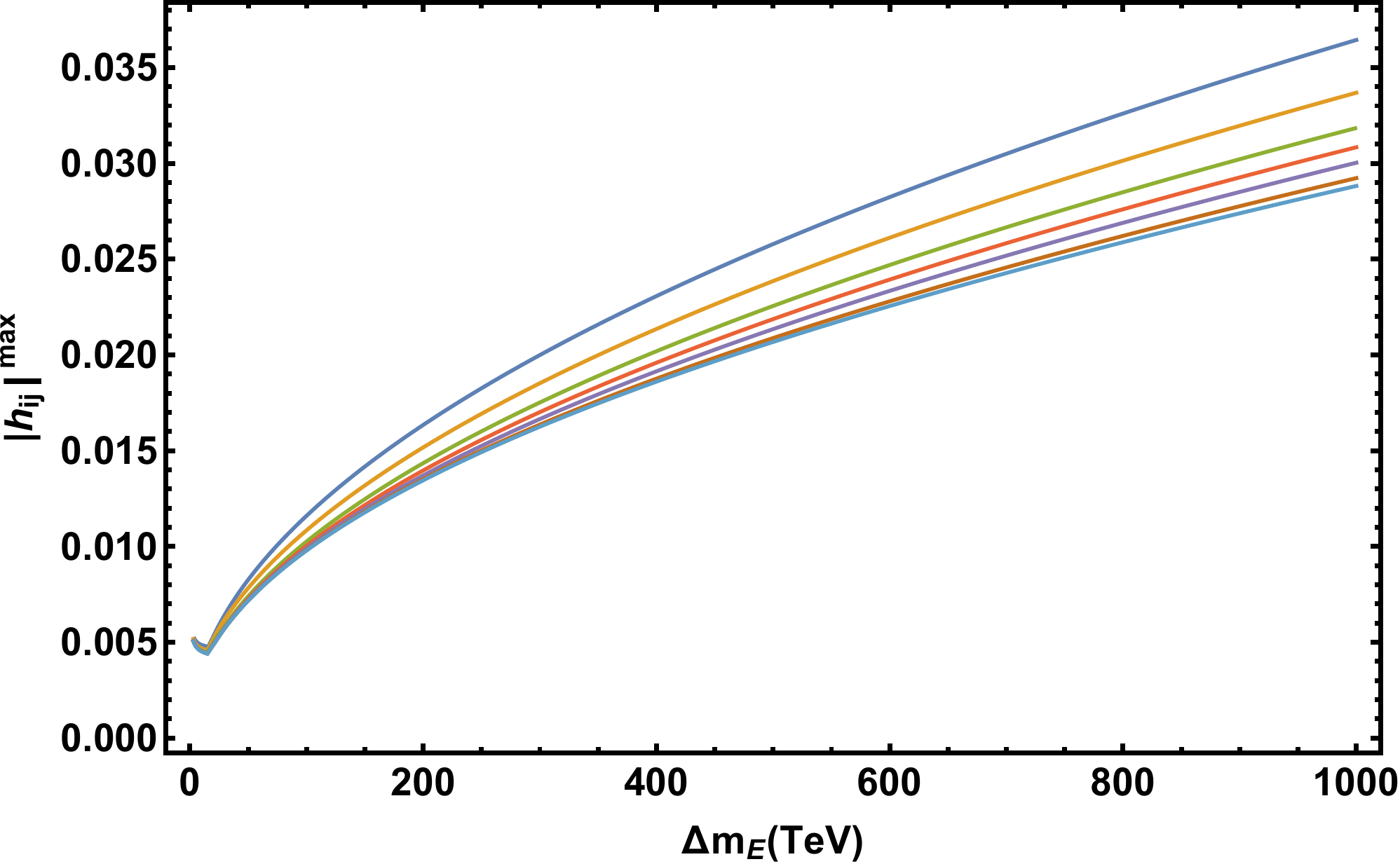}
\includegraphics[width=1. in]{legends.pdf}
\caption{Numerical solutions for $2\times 2$ mixing simultaneously fitting Eqs. (\ref{eq:cip3}) and (\ref{eq:cip4}) for the case of inverted ordering of neutrino masses: (top left) $SU(2)_R$ breaking scale $u_R$,  (top right) $2\times 2$ PMNS mixing angle, (middle left) the arbitrary rotation matrix angles in ${\mathcal{R}_l}$, (middle right) the arbitrary rotation matrix angles in ${\mathcal{R}_\nu}$, (bottom) the maximal element of the Yukawa matrix ${\bf h}$, as a function of the mass difference between two generations of vector-like charged lepton masses $\Delta m_E$ for different benchmark values of $m_B$. See text for the benchmark values of the other relevant parameters.}
\end{figure}

In Fig. 2, we present the numerical solutions for the case of normal ordering, setting the lightest neutrino mass to be $10^{-2}$ eV as a benchmark choice and using the resultant heavier neutrino masses as the diagonal entries of $m^\text{diag}_\nu$. In the top left plot we show the relevant $SU(2)_R$ breaking VEV $u_R$ as a function of the mass difference between two generations of vector-like charged lepton masses $\Delta m_E$ for different benchmark values of $m_B$ and in the top right plot we show the variation of the $2\times 2$ PMNS mixing angle corresponding to the usual mixing angle $\theta_{23}$ as a function of the mass difference between two generations of vector-like charged lepton masses $\Delta m_E$ for different benchmark values of $m_B$. Vector-like quark masses are limited to be heavier than $m_Q \gtrsim 1.4~\text{TeV}$ \cite{Sirunyan:2018qau,Aaboud:2018ifs,Aaboud:2018pii}, which is satisfied for all our choices. 

It is evident from these plots that one requires a $SU(2)_R$ breaking VEV of $u_R\sim \mathcal{O} (10^2)$~TeV to generate the correct neutrino mass splitting and a maximal PMNS mixing angle. Although, \textit{a priori}, it appears to be relatively high as compared to the currently accessible mass scales at the LHC, it is interesting to note that such mass scales are in agreement with the strong cosmological bounds (discussed in Section \ref{sec:pheno}) on the $SU(2)_R$ breaking scale in this model. As mentioned earlier, these plots also clearly demonstrate that the hierarchy of masses of the two generations of vector-like charged lepton masses play a crucial role in generating a non-trivial PMNS mixing angle in contrast to the scenario with a single generation of vector-like charged leptons or more than one generation of vector-like charged leptons with degenerate masses where the neutrino mass matrix turns out to be directly proportional to the charged lepton mass matrix leading to a trivial PMNS mixing matrix which is inconsistent with the neutrino oscillation data. A large splitting $\Delta m_{E}\gtrsim 100~\text{TeV}$ is thus required to achieve a large neutrino mixing angle. In our benchmark choice, this is achieved using a hierarchical heavy fermion spectrum. Note that only a strictly hierarchical spectrum with $\Delta m_{E}/m_{E_1}\gtrsim 100$ can lead to the maximal neutrino mixing angle case. In the middle two plots we show the arbitrary rotation matrix angles in ${\mathcal{R}_l}$ and ${\mathcal{R}_\nu}$ defined in Eq. (\ref{eq:na1}) as a function of the mass difference between two generations of vector-like charged lepton masses $\Delta m_E$ for different benchmark values of $m_B$. Finally, in the bottom plot we show the maximal element of the Yukawa matrix ${\bf h}$ as a function of the mass difference between two generations of vector-like charged lepton masses $\Delta m_E$ for different benchmark values of $m_B$, which shows that the numerical solutions correspond to Yukawa couplings well within the perturbative regime. 

In Fig. 3, we present the numerical solutions for the case of inverted ordering using the best-fit values for the atmospheric and solar neutrino mass squared differences from the global oscillation analysis of~\cite{deSalas:2017kay}, setting the lightest neutrino mass to be $10^{-2}$ eV as a benchmark choice and using the second and third generation neutrino masses as the diagonal entries of $m^\text{diag}_\nu$. We note that in this case one also requires a $SU(2)_R$ breaking VEV of $u_R\sim \mathcal{O} (10^2)$~TeV to generate the correct neutrino mass splitting and a maximal PMNS mixing angle. 

%%%%%%%%%%%%%%%%%%%%%%%%%%%%%%%%%%%%%%%%%%%%%%%%%%%%%%%%%%%%%%%
\section{Left-Right Symmetric Model with a global $B-L$ symmetry in the presence of a bi-doublet Higgs}{\label{app:0}}
In this alternative scenario a Higgs bi-doublet breaks the electroweak symmetry. The field content and their transformations are summarised in Table. \ref{tab:LR3}. The quarks acquire their masses through the vacuum expectation value of the bi-doublet while the Yukawa couplings giving rise to lepton masses are forbidden by some symmetry\footnote{For example one may introduce an additional discrete $Z_2$ symmetry, such that $L_R$, $\sigma_R$ and $E_R$ are odd under this discrete symmetry. Note that in such a case the vector-like mass term for $\sigma$ and $E$  will break this $Z_2$ symmetry softly.}. Both the charged 
and neutral leptons would then acquire Dirac seesaw masses in this scenario \cite{fer-mass}\footnote{For other interesting implementations of purely Dirac neutrino masses in the context of other models see for example \cite{Chulia:2016ngi, CentellesChulia:2018gwr, CentellesChulia:2018bkz, Bonilla:2018ynb}.}. For this 
purpose, we introduce four singlet vector-like fermions, which are the charged and neutral heavy leptons:
\begin{eqnarray}
\sigma_L  \equiv [ 1,1,1, 0 ]\,,\quad
\sigma_R  \equiv [ 1,1,1, 0 ]\,,\quad
E_L  \equiv [ 1,1,1, 2 ]\,, \quad
E_R  \equiv [ 1,1,1, 2 ]\,,
\end{eqnarray}
which carry $B-L=1$, and hence, $\zeta =-1$ for the neutral fermions $\sigma_{L,R}$ and $\zeta =1$ for the charged 
fermions $E_{L,R}$. The left-right symmetry breaking will allow mixing of these fermions with the light 
leptons, and hence, the assignment of lepton number is more natural than the conventional left-right 
symmetric models, where similar new singlets carry vanishing lepton numbers. The VEVs of the fields 
$\chi_{L,R}$ introduce mixing of the new neutral leptons $\sigma_{L,R}$ with the neutrinos and the new 
charged leptons $E_{L,R}$ with the charged leptons. As far as quark masses are concerned, vector-like 
heavy quark fields are not necessary for this scheme, but nonetheless can be included.
\begin{table}[t]
%[t!]
\begin{center}
\begin{tabular}{|c|c|c|c|c|c|c|}
\hline
Field     & $ SU(2)_L$ & $SU(2)_R$ & $B-L$ & $\zeta$ & $X=(B-L) +\zeta$ & $SU(3)_C$ \\
\hline
$q_L$        &  \bf{2}         & \bf{1}    & 1/3  & 0      & \bf{1/3}   & \bf{3}   \\
$q_R$        &  \bf{1}         & \bf{2}    & 1/3  & 0      & \bf{1/3}   &\bf{3}   \\
$\ell_L$     &  \bf{2}        & \bf{1}    & $-1$    & 0     & $\mathbf{-1}$    & \bf{1}   \\
$\ell_R$     &  \bf{1}         & \bf{2}    & $-1$    & 0     & $\mathbf{-1}$    & \bf{1}   \\
\hline
$E_{L,R}$ &  \bf{1}         & \bf{1}    & $-1$   & $-1$     & $\mathbf{-2}$    & \bf{1}   \\
$\sigma_{L,R}$ &  \bf{1}         & \bf{1}    & $-1$   & +1     & \bf{0}     & \bf{1}   \\
\hline
 $\chi_L$   &  \bf{2}        & \bf{1}    & 0    & 1       & \bf{1}    & \bf{1}   \\
 $\chi_R$   &  \bf{1}       & \bf{2}    & 0    & 1       & \bf{1}    & \bf{1}   \\
 $\rho$     &  \bf{1}        & \bf{1}    & 0    & 0       & \bf{0}    & \bf{1}   \\
 $\Phi$   & \bf{2}         & \bf{2}    & 0    & 0       & \bf{0}    & \bf{1}   \\
\hline
\end{tabular}
\end{center}
\caption{Field content of the LRSM with a unbroken $B-L$ symmetry in the presence of a Higgs bi-doublet.}
\label{tab:LR3}
\end{table}
The general scalar potential with all the scalar fields can be written as
\begin{eqnarray}
\lefteqn{V= \mu_1^2 \text{Tr}[\Phi^{\dagger}\Phi]+\mu_2^2 (\text{Tr}[\tilde{\Phi} \Phi^{\dagger}]+
\text{Tr}[\tilde{\Phi}^{\dagger}\Phi] )+ \lambda_1 (\text{Tr}[\Phi^{\dagger} \Phi])^2+
\lambda_2[(\text{Tr}[\tilde{\Phi} \Phi^{\dagger}])^2+(\text{Tr}[\tilde{\Phi}^{\dagger}\Phi ])^2 ]}
\nonumber \\
&~~~~~~~ +\lambda_3 \text{Tr}[\tilde{\Phi} \Phi^{\dagger}]\text{Tr}[\tilde{\Phi}^{\dagger}\Phi ]+
\lambda_4 \text{Tr}[\Phi^{\dagger}\Phi](\text{Tr}[\tilde{\Phi} \Phi^{\dagger}]+
\text{Tr}[\tilde{\Phi}^{\dagger}\Phi] )+\mu^2_h (\chi^{\dagger}_L \chi_L+\chi^{\dagger}_R \chi_R)
\nonumber \\
&~~~~~~ + \lambda_5 [(\chi^{\dagger}_L \chi_L)^2+(\chi^{\dagger}_R \chi_R)^2]+\lambda_6 (\chi^{\dagger}_L \chi_L)
(\chi^{\dagger}_R \chi_R)+\alpha_1 \text{Tr}[\Phi^{\dagger}\Phi](\chi^{\dagger}_L \chi_L+ \chi^{\dagger}_R \chi_R)
\nonumber \\
&~~~~~~~ +\alpha_2 (\chi^{\dagger}_L\Phi\Phi^{\dagger} \chi_L+\chi^{\dagger}_R\Phi^{\dagger}\Phi \chi_R)+
\alpha_3 (\chi^{\dagger}_L\Phi_2\Phi_2^{\dagger} \chi_L+\chi^{\dagger}_R\tilde{\Phi}^{\dagger}\tilde{\Phi} \chi_R)+
\alpha_4 (\chi^{\dagger}_L\Phi\tilde{\Phi}^{\dagger} \chi_L
\nonumber \\
& +\chi^{\dagger}_R\Phi^{\dagger}\tilde{\Phi} \chi_R)+\alpha^*_4 (\chi^{\dagger}_L\tilde{\Phi}\Phi^{\dagger} \chi_L+
\chi^{\dagger}_R\tilde{\Phi}^{\dagger}\tilde{\Phi} \chi_R) + \mu_{h\Phi1} (\chi^{\dagger}_L \Phi \chi_R + \chi^{\dagger}_R
\Phi^{\dagger} \chi_L)
\nonumber \\
& \hspace{-2.6em}+\mu_{h\Phi2} (\chi^{\dagger}_L \tilde{\Phi} \chi_R+ \chi^{\dagger}_R \tilde{\Phi}^{\dagger} \chi_L)-\mu^2_{\rho} \rho^2 +\lambda_7 \rho^4 +
M \rho(\chi^{\dagger}_L \chi_L -\chi^{\dagger}_R \chi_R)
\nonumber \\
& \hspace{-1.7em}+\lambda_8 \rho^2(\chi^{\dagger}_L \chi_L+
\chi^{\dagger}_R \chi_R)+ \lambda_9 \rho^2 \text{Tr}[\Phi^{\dagger}\Phi]+\lambda_{10}\rho^2
[\text{Det}(\Phi)+\text{Det}(\Phi^{\dagger})]\,,
\end{eqnarray}
where $ \tilde{\Phi} = \tau_2 \Phi^* \tau_2 $. Using the notation $\langle \chi_L\rangle =u_L$, $\langle \chi_R\rangle =u_R$, $\langle \Phi\rangle = \text{diag}(v_1,v_2)$ and $\langle \rho \rangle =s$, we minimise the scalar potential to obtain
\begin{eqnarray}
\mu^2_L u_L + 2 \lambda_5 u^3_L+
\lambda_6 u_L u^2_R+ \mu_{h\phi} (v_1+v_2)u_R = 0
\label{eq1}\,,  \\
\mu^2_R u_R + 2 \lambda_5 u^3_R+
\lambda_6 u_R u^2_L +\mu_{h\phi} (v_1+v_2)u_L = 0
\label{eq2}\,,
\end{eqnarray}
where $\mu_{h\phi}=(\mu_{h\Phi1} v_2 + \mu_{h\Phi2} v_1)/(v_1+v_2)$. The effective mass terms $\mu^2_L$ and $\mu^2_R$ are given by
\begin{eqnarray}
\mu^2_L = \mu^2_h +M s + \lambda_8 s^2 + (\alpha_4+\alpha^*_4) v_1v_2+ \alpha_1
(v^2_1+v^2_2)+\alpha_2 v^2_2 +\alpha_3 v^2_1  \nonumber \,, \\
\mu^2_R = \mu^2_h -M s + \lambda_8 s^2 + (\alpha_4+\alpha^*_4) v_1v_2+ \alpha_1
(v^2_1+v^2_2)+\alpha_2 v^2_2 +\alpha_3 v^2_1\,.
\label{mr}
\end{eqnarray}
From Eqs. (\ref{eq1}), (\ref{eq2}) one gets
\begin{equation}
u_Lu_R (2 M s) + (2 \lambda_5 -\lambda_6)(u^2_L-u^2_R)u_Lu_R+\mu_{h\phi} (v_1+v_2)(u^2_R-u^2_L) = 0\,.
\end{equation}
One can derive the seesaw relation from the above equation as
\begin{equation}
u_Lu_R = \frac{\mu_{h\phi}(v_1+v_2)(u^2_L-u^2_R)}{2M s + (2\lambda_5-\lambda_6)(u^2_L-u^2_R)}\,.
\end{equation}
Assuming the hierarchy $u_L \ll u_R \ll s, M $ yields
\begin{equation}
u_L = \frac{-\mu_{h\phi} (v_1+v_2) u_R}{2 M s}\,.
\end{equation}
Thus in this scenario a small $u_L/u_R $ can be obtained by choosing the scales
$M, \rho, \mu_{h\phi}$ appropriately.

The Yukawa term for the quarks involving the Higgs bi-doublet is given by
\begin{eqnarray}
    -{\cal L}_{\text{bi-doublet}} &=& f_{ij} \bar{q}_L q_{R} \Phi + f'_{ij} \bar{q}_L q_{R} \tilde{\Phi}  +{ \rm{h.c.}} \; ,
\end{eqnarray}
where $\tilde{\Phi}=\tau_2 \Phi \tau_2$ and $\tau_2$ is the second Pauli matrix. After the electroweak symmetry is broken via the VEV of the Higgs scalar bi-doublet, one can obtain the Dirac mass terms for the SM quarks. Thus, the quark masses are similar to those in the conventional LRSM and we will not repeat the details here\footnote{Note that, in the presence of vector-like quarks in the model there can be a seesaw type contribution as well \cite{Deppisch:2016scs}.}. 
On the other hand, for the charged and neutral leptons there is no Dirac mass term due to the Higgs bi-doublet as mentioned earlier\footnote{To ensure this we assume that under a discrete $Z_2$ symmetry the right-handed fields $L_R$, $\sigma_R$, and $E_R$ are odd, while all other fields are even.}. 
The Yukawa interactions giving mass to the leptons are given by 
\begin{eqnarray}
    {\cal L} &=& 
    f_L L_L^T C^{-1} \sigma_L \chi_L + f_R L_R^T C^{-1} \sigma_R  \chi_R + m_\sigma \overline{\sigma_L} ~\sigma_R+
 h_L \bar{L}_L  \chi_L E_R \nonumber
 \\
 &&+ ~h_R \bar{L}_R  \chi_R E_L
    + m_E \overline{E_L} ~E_R +\text{h.c.}\,.
\end{eqnarray}
The charged lepton masses are generated through a Dirac seesaw mechanism (similar to Section \ref{sec2}) and the mass matrix is given by
\begin{equation}{\label{lepton:mass}}
    m_{l_{ij}} = u_L u_R h_{L_{ik}} M^{-1}_{E_{k}} h^{\dagger}_{R_{kj}}\,.
\end{equation}
To simplify the analysis of the neutrino sector we shall work with the CP conjugates of the right-handed
fields
\begin{equation} \nu_R \stackrel{\mathrm{CP}}{\rightarrow} {(\nu_R)}^c = {(\nu^c)}_L = N_L
~~~ {\rm and} ~~~ \sigma_R \stackrel{\mathrm{CP}}{\rightarrow} {(\Sigma^c)}_L = \Sigma_L \,,
\end{equation}
so that the neutrino mass matrix can be written in the basis
$\pmatrix{ \nu_L & N_L & \sigma_L & \Sigma_L}$ as
\begin{equation}
    {\cal M}_\nu = \pmatrix{ 0 & 0 &a & 0 \cr 0 & 0 & 0 & b \cr
    a & 0 & 0 & c \cr 0 & b & c & 0} \,.
\end{equation}
Here $a = f_L u_L$; $b = f_R u_R$; 
and $c = m_\sigma$. This gives six Dirac neutrinos, three very heavy ones with mass $\sim c$, and three 
light ones with mass $\sim ab/c$ \cite{dn1,dn2,dn3}. The heavy Dirac neutrinos are made of $\sigma_L$ and $\Sigma_L$, while the light Dirac neutrinos are the usual neutrinos, a combination of $\nu_L$ and $N_L$ or $\nu_R$. Note that one can \textit{a priori} draw a two loop diagram similar to Fig. 1, without the vector-like fields in a scenario where the charged lepton and quark masses are generated by the bi-doublet Higgs and only neutrino masses are vanishing at the tree level. However, in such a diagram the external neutrino lines can be folded to generate a tadpole correction to the VEV of the neutral component of bi-doublet Higgs which diverges and therefore must be cancelled by adding a counterterm \cite{Branco:1978bz}. Therefore, one must ensure that the bi-doublet VEV satisfies the constraint $\langle 0 |\mathrm{Tr}[\Phi \tau_2 \Phi^{\ast}\tau_2] |0\rangle=0$ at the tree level, implying that there is no mixing between $W_L-W_R$ at the tree level.

%%%%%%%%%%%%%%%%%%%%%%%%%%%%%%%%%%%%%%%%%%%%%%%%%%%%%%
\section{Phenomenology and constraints}{\label{sec:pheno}}
We now briefly outline the general observable phenomenology of our LRSM, specifically the complimentary constraints cosmology and direct collider searches can put on additional gauge bosons to the SM. These constraints can be interpreted in the $M_{Z'}-g'$ parameter space of an unbroken additional gauge group $U(1)_X$, where $M_{Z'}$ is the mass of the $U(1)_X$ mediator ($Z'$) and $g'$ is the coupling strength of $Z'$ to fermions. They are however directly transferable to the $M_{W_R}-g_R$ parameter space of our model. Given the benchmark parameter values considered in this paper, and the subsequent $\sim \mathcal{O}(10^2) ~\text{TeV}$ size of the $SU(2)_R$ breaking scale, we are most interested in constraints in the region $M_{W_R} > 1 ~\text{TeV}$.

The bound on the number of fermionic relativistic degrees of freedom at the time of Big Bang Nucleosynthesis (BBN), $N_{\mathrm{eff}}< 4$ (obtained at 90\% CL from the abundances of light nuclei), can exclude an important region in the generic $M_{Z'}-g'$ parameter space. With the addition of right-handed neutrinos $\nu_R$ to the SM, the $U(1)_X$ mediator can lead to the thermalisation of $\nu_{R}$ with the photon bath via the process $\bar{f}f\leftrightarrow \bar{\nu}_R \nu_R$. In particular, the size of this effect can be increased through resonant enhancement at temperatures around $M_{Z'}$ when the mediator goes on-shell. 

Over the mass range $ 1~\mathrm{eV} < M_{Z'} < 1~\mathrm{TeV} $, BBN puts a varying upper bound on the coupling $g'$ from the condition that the thermalisation of $\nu_{R}$ does not contribute considerably to $N_{\mathrm{eff}}$. For masses $M_{Z'} < 1 ~\mathrm{MeV}$, for example, $\nu_R $ must thermalise after the photon temperature $\sim 1~\mathrm{MeV}$, giving $g' < 3\times10^{-7}~\mathrm{keV}/M_{Z'}$. Natural couplings of order unity are similarly excluded for $1 ~\mathrm{MeV} < M_{Z'} < 10 ~\mathrm{GeV}$, but both of these regimes are clearly not of interest in our scenario. For $M_{Z'} > 10 ~\mathrm{GeV}$ the $\bar{f}f\leftrightarrow \bar{\nu}_R \nu_R$ process can be treated at $T_{\mathrm{BBN}}$ as a four-fermion contact interaction and constraints are thus put on the ratio $M_{Z'}/g'$. This is analogous to a constraint on the ratio $M_{W_R}/g_{R}$, which also leads to thermalisation via $\bar{f}f\leftrightarrow \bar{\nu}_R \nu_R$. The constraint presented in Ref.~\cite{Barger:2003zh} is $M_{Z'}/g' > 6.7 ~\mathrm{TeV}$. For the benchmark couplings considered in the radiative and Higgs bi-doublet cases in this paper, this puts a lower bound on $M_{W_R}$ in the range $1 - 6~\mathrm{TeV}$. Reference~\cite{Anchordoqui:2012qu} similarly investigated the regime $M_{Z'} \gg T_{\text{BBN}}$, but instead studied the relationship between $N_{\mathrm{eff}}$ and the temperature $T_{\nu_R}^{\mathrm{dec}}$ at which $\nu_R$ decouples. Enforcing the interaction rate $\Gamma(T_{\nu_R}^{\mathrm{dec}}) $ to be equal to the Hubble rate $H(T_{\nu_R}^{\mathrm{dec}})$ at this temperature, a bound of similar size can be placed on $M_{W_R}/g_R$.

Direct searches for additional gauge bosons have also been performed at colliders, with analyses probing large values of $M_{Z'}$. The study of LEP~2 data \cite{Schael:2013ita} in Ref. \cite{Heeck:2014zfa} parametrises the effect of $Z'$ exchange on di-electron and di-muon channels with a four-fermion contact interaction for $ M_{Z'} \gg 200 ~\mathrm{GeV}$. This improves on similar model independent bounds from the CDF and D\O{} experiments at the Tevatron \cite{Abulencia:2005nf,Carena:2004xs} to $M_{Z'}/g' > 6.9~\mathrm{TeV}$. In the mass range $M_{Z'} = 0.5 - 3.5 ~\mathrm{TeV}$, the ATLAS and CMS experiments at the LHC constrained slightly more of the parameter space than the linear constraint on $M_{Z'}/g' $ \cite{Aad:2014cka,Khachatryan:2014fba}. A more recent ATLAS analysis set a lower bound on the mediator mass of $ M_{Z'} > 5.1 \, \text{TeV}$ using the Sequential Standard Model (SSM) benchmark scenario, where the couplings $g'$ are the same as those of the SM \cite{Aad:2019fac}. For the benchmark value of $g_R$ considered in this paper and the subsequent lower bound of $M_{W_R}\gtrsim 5 ~\text{TeV}$, we can safely expect the additional gauge boson $W_R$ to be out of reach at the high-luminosity LHC \footnote{For a relevant discussion of $W_R$ multi-leptonic decay modes see \cite{Das:2017hmg, Das:2016akd}.}.

\section{Conclusion}
\label{sec:conclusion}
The question of how neutrinos acquire their masses, which are needed to understand the observed oscillation phenomena, remains one of the main outstanding issues in particle physics. The overwhelming majority of explanations work by generating $\Delta L = 2$ Majorana masses for neutrinos, with the type-I seesaw mechanism as the most prominent example. While this approach clearly has theoretical and phenomenological advantages, it is also important to pursue other potential solutions. 

In this paper, we have proposed an alternative formulation of a Left-Right Symmetric Model where $B-L$ is not broken and thus neutrino Majorana masses are strictly forbidden. Instead, $B-L$ remains a global symmetry after the left-right symmetry breaking, allowing only Dirac mass terms for neutrinos. While parity is restored at a high scale, this formulation provides a natural framework to explain $B-L$ as an anomaly-free global symmetry of the SM. In this model, a bi-doublet Higgs is not present and the charged SM fermion masses fundamentally originate from a Dirac seesaw mechanism connected to heavy vector-like fermion partners. The lightness of neutrinos in the instance is explained as neutrino Dirac mass terms are induced at the two-loop level, cf. Fig.~1. Alternatively, a Dirac seesaw mechanism can be invoked for the neutrinos as well if the corresponding heavy vector-like neutrino partners exist. We showed that for an appropriate spectrum of heavy states, both the lightness of neutrinos relative to the charged fermions and a large two flavour mixing in the leptonic sector can be explained. An analysis of the full three-flavour framework will be reported elsewhere.

Our models may be enhanced in several directions. For example, while neutrinoless double beta decay is strictly forbidden, one can add a light charge-neutral scalar particle $\phi$ with quantum numbers $[1,1,2,1]$ under our model gauge group. This particle can potentially be a Dark Matter candidate \cite{Berezinsky:1993fm, Garcia-Cely:2017oco, Brune:2018sab} with a Yukawa coupling to the heavy $N$ of the form $g_\phi N N \phi$. In this case, $0\nu\beta\beta\phi$ decay with emission of the light neutral scalar $\phi$ via a single effective dimension-7 operator of the form $\Lambda_\text{NP}^{-3}(\bar u\mathcal{O} d)(\bar e\mathcal{O}\nu)\phi$ is possible. This provides a working example of a scenario where purely Dirac neutrinos can mimic the conventional $0\nu\beta\beta$ decay associated with the violation of lepton number by two units and thus the Majorana nature of neutrinos. This supports the necessity of searches for extra particles in double beta decay in order to fully understand the nature of neutrinos \cite{Cepedello:2018zvr}.

Finally, we would also like to make some remarks on the possibility of realising leptogenesis in this formalism. In our scenarios, leptogenesis may occur through neutrinogenesis \cite{dl1, dl2}. To give an example, the scalar field $\chi_R$ can decay as $\chi_R \to \ell_R + E_R$ and $\chi_R \to \Phi^\dagger + \Phi$ because of the coupling $\chi_R^\dagger \chi_R \Phi^\dagger \Phi$ when $\chi_R$ acquires a VEV. Through self-energy diagrams there can then be an interference and these decays can generate an asymmetry in the $\zeta$ quantum number which means that there will be more $E_R$ compared to $E_L$, since $\ell_R$ and $\phi$ have $\zeta = 0$. However, since $B-L$ is conserved, the asymmetry in $E_R$ will be the same as the asymmetry in $\ell_R$. Since $B-L$ is conserved, the out-of-equilibrium three-body decays of $E_R$ and $E_L$ will produce different amounts of $\nu_L$ and $\nu_R$. Since the Yukawa couplings responsible for $\nu_R + \phi \to \nu_L + W_L$ are not allowed, the amount of lepton asymmetry stored in $\nu_R$ will not be converted into $\nu_L$. Thus although there is no $B-L$ asymmetry, there is an asymmetry in $\nu_L$ and an equal and opposite amount of asymmetry in $\nu_R$. Since the $\nu_R$ asymmetry will not get converted to a baryon asymmetry in the presence of sphalerons, the $\nu_L$ asymmetry will generate the baryon asymmetry of the universe. Since $B-L$ is an unbroken symmetry in this model, there are no other washout interactions that can affect the baryon asymmetry of the universe. Alternatively, one can also add an additional heavy doublet scalar field to implement a neutrinogenesis mechanism similar to Ref.~\cite{Gu:2007mc}.

\begin{acknowledgments}
PDB and FFD acknowledge support from the Science and Technology Facilities Council (STFC). CH acknowledges support within the framework of the European Unions Horizon
2020 research and innovation programme under the Marie Sklodowska-Curie grant agreements No 690575 and No 674896. US acknowledges support from the JC Bose National Fellowship grant under DST, India. 
\end{acknowledgments}

\appendix

\section{Evaluation of the two loop integral using the Passarino-Veltman integral reduction}{\label{app:A}}
In this appendix we outline the evaluation of the two loop integral given in Eq. (\ref{eq:integral}) using the Passarino-Veltman integral reduction. Note that the first term in the numerator of Eq. (\ref{eq:integral}) is suppressed by $m_{W_L}^2/m_{W_R}^2$ with respect to the second term and therefore can be neglected to obtain
\begin{eqnarray}{\label{eq:appA1}}
\mathcal{I}_k\simeq \int \frac{d^{4}k}{\left(2\pi\right)^{4}}\,
\int \frac{d^{4}p}{\left(2\pi\right)^{4}}\,
\frac{1}{p^{2} (p^{2}-m_{E_{k}}^2) k^2 (k^2-m_B^2) (p+k)^2 [(p+k)^2-m_{T}^2]}\, .
\end{eqnarray}
Next using Partial-fraction decomposition and Passarino-Veltman reduction formula the integral can be simplified to obtain
\begin{eqnarray}{\label{eq:appA2}}
\mathcal{I}_k=\frac{i}{16 \pi^2 m_T^2 m_B^2 m_{E_k}^2} \int \frac{d^{4}k}{\left(2\pi\right)^{4}} \left[\frac{1}{k^2-m_B^2}-\frac{1}{k^2}\right]\,
&&\left[ B_0(k^2,m_E^2,m_T^2) -B_0(k^2,0,m_T^2)\right.\nonumber\\
&&\left.-B_0(k^2,m_E^2,0) +B_0(k^2,0,0)  \right]\, ,
\end{eqnarray}
where $B_{0}$ is the Passarino-Veltman function defined
as~\cite{Passarino:1978jh} 
\begin{equation}{\label{eq:appA3}}
B_{0}\left(k^{2},m_{1}^{2},m_{2}^{2}\right)  \,=\, 
\frac{1}{\epsilon}-
\int_{0}^{1} dx \ln \left(
\frac{-x\,\left(1-x\right)\,k^{2}\,+\,\left(1-x\right)\,m_{1}^{2}
\,+\,x\,m_{2}^{2}}{\mu^{2}}\right)\,.
\end{equation} 
Next performing a Wick rotation and defining the dimensionless quantities $\alpha_k=m_B^2/m_{E_k}^2$ and $\beta_k=m_T^2/m_{E_k}^2$ the integral given in Eq. (\ref{eq:appA2}) can be further simplified to obtain
\begin{eqnarray}{\label{eq:appA4}}
\mathcal{I}_k= \frac{1}{(16 \pi^2)^2 m_B^2 m_T^2}\int^{\infty}_{0} \,dr \, \frac{\alpha_k}{r+\alpha_k}\,
\int ^{1}_{0}\, dx\, \ln\left[ 
\frac{x(1-x)r+(1-x)}{x(1-x)r+(1-x)+x\beta_k} \,\frac{(1-x)r+\beta_k }{(1-x)r}
\right]\, .\nonumber\\
\end{eqnarray}
%%%%%%%%%%%%%%%%%%%%%%%%%%%%%%%%%%%%%%%%%%%%%%%%%%%%%%
\section{Evaluation of the two loop integral using master integral reduction}{\label{app:B}}
In this section we outline another alternative approach using master integral reduction for the evaluation of the two loop integral given in Eq. (\ref{eq:appA1}). Using Feynman parametrisation the two loop integral can be written as
\begin{eqnarray}{\label{nintegral1}}
\mathcal{I}_k=  \int _{0}^{1}  \int _{0}^{1}\int _{0}^{1} dx_1  dx_2   dx_3  G(m_1(x_1, m_{E_k}),2; m_2(x_2 , m_B),2; m_3(x_3, m_T),2;0)\,,
\end{eqnarray}
where
\begin{eqnarray}{\label{nintegral2}}
G(m_1,\alpha_1;m_2,\alpha_2;m_3,\alpha_3;q^2)=\int {d^D}p\; {d^D}k  \frac{1}{(p^2-m_1^2)^{\alpha_1}(k^2-m_2^2)^{\alpha_2}[(p+k+q)^2-m_3^2)]^{\alpha_3}}\,.\nonumber\\
\end{eqnarray}
The integration given by Eq. (\ref{nintegral2}) can be obtained by taking derivative of the basic master integral
\begin{eqnarray}{\label{nintegral3}}
G(m_1,2;m_2,2;m_3,2;0)=\partial_{m_2^2}\partial_{m_3^2} G(m_1,2;m_2,1;m_3,1;0)\,,
\end{eqnarray} 
where the master integral is given by \cite{Ghinculov:1994sd}
\begin{eqnarray}
G(m_1,2;m_2,1;m_3,1;0) = 
   &&~\pi^{4} \Big[   \frac{2}{\epsilon^{2}}
    + \frac{1}{\epsilon} [- 1 + 2 \gamma + 2 \log (\pi \, m_{1}^{2}) ]
    + \frac{1}{4} + \frac{\pi^{2}}{12}
                                                 \nonumber\\ && ~+ \frac{1}{4} [- 1 + 2 \gamma
                                                + 2 \log (\pi \, m_{1}^{2}) ]^{2} - 1 + g(m_{1},m_{2},m_{3};0)\Big]\,,
\end{eqnarray}
with 
\begin{equation}{\label{nintegral4}}
g(m_{1},m_{2},m_{3};0) = \int_{0}^{1} dx \,
 [ \, 1 + Sp(1-\mu^{2}) - \frac{\mu^{2}}{1-\mu^{2}} \log \mu^{2} \, ]
      \; ,
\end{equation}
where $Sp(z)$ corresponds to the Spence function and the following notations are used
\begin{eqnarray}{\label{nintegral5}}
   \mu^{2}  & = &  \frac{a x + b (1-x)}{x (1-x)}\, ,  \quad
         a   =   \frac{m_{2}^{2}}{m_{1}^{2}} \, , \quad
         b = \frac{m_{3}^{2}}{m_{1}^{2}}\,.
\end{eqnarray}

\end{document}